\documentclass[journal]{IEEEtran}
\usepackage{amsmath,graphicx}
\usepackage{multirow}
\usepackage{multicol}
\usepackage{booktabs}
\usepackage[table]{xcolor}
\usepackage{caption}
\usepackage{amsfonts}
\usepackage[numbers]{natbib}
\usepackage{subcaption}
\usepackage[flushmargin]{footmisc}
\usepackage[numbers]{natbib}
\usepackage{url}
\usepackage[%
breaklinks=true,
colorlinks=true, 
citecolor=black,
urlcolor=black,
linkcolor=black,
linktocpage,
draft=false,
pageanchor=true,
hyperfigures=true,
plainpages=false
]{hyperref}
\usepackage{tikz}

\title{Phone Duration Modeling for Speaker Age Estimation in Children}
\author{Prashanth Gurunath Shivakumar,~\IEEEmembership{Student Member,~IEEE,}
        Somer~Bishop,~\IEEEmembership{} \\
        Catherine Lord,~\IEEEmembership{}
        Shrikanth Narayanan,~\IEEEmembership{Fellow,~IEEE}%
\thanks{P. Shivakumar and S. Narayanan are with the Department of Electrical and Computer Engineering, University of Southern California, Los Angeles, California 90089, USA (e-mail: pgurunat@usc.edu;shri@sipi.usc.edu). S. Bishop is with Department of Psychiatry, University of California, San
Francisco, USA e-mail:(somer.bishop@ucsf.edu).
C. Lord is with the Semel Institute of Neuroscience and Human Behavior, University of California, Los Angeles, USA e-mail:(CLord@mednet.ucla.edu).}}
\begin{document}
%
\maketitle
\begin{abstract}
Automatic inference of important paralinguistic information such as age from speech is an important area of research with numerous spoken language technology based applications.
Speaker age estimation has applications in enabling personalization and age-appropriate curation of information and content.
However, research in speaker age estimation in children is especially challenging due to paucity of relevant speech data representing the developmental spectrum, and the high signal variability especially intra age variability that complicates modeling.
Most approaches in children speaker age estimation adopt methods directly from research on adult speech processing.
In this paper, we propose features specific to children and focus on  speaker's phone duration as an important biomarker of children's age.
We propose phone duration modeling for predicting age from child's speech.
To enable that, children speech is first forced aligned with the corresponding transcription to derive phone duration distributions.
Statistical functionals are computed from phone duration distributions for each phoneme which are in turn used to train regression models to predict speaker age.
Two children speech datasets are employed to demonstrate the robustness of phone duration features.
We perform age regression experiments on age categories ranging from children studying in kindergarten to grade 10.
Experimental results suggest phone durations contain important development-related information of children.
Phonemes contributing most to estimation of children speaker age are analyzed and presented.

\end{abstract}
\section{Introduction}
\label{sec:intro}
\IEEEPARstart{S}{peech} contains important paralinguistic information including speaker's age, gender, emotions, and other behavior constructs \cite{Schuller2013ParalinguisticsinSpeechand}.
Knowledge of such information can help improve spoken language technologies (SLT) such as speech recognition, speaker recognition, and enhance the experience of SLT based applications by providing robustness against variability along those dimensions.
Inference of age, gender from children speech can help better tailor conversational interfaces such as education and learning platforms, entertainment, interactive gaming, tutoring, social networking for different age-gender demographics. Speech-based biomarkers are also increasingly used in supporting health applications\cite{Bone2017SignalProcessingandMachine}, including related to developmental disorders \cite{bone2017behavioral}.

Knowledge of an individual's age is an important meta-data for several applications.
Automatic recognition of speaker age is valuable especially when speech is the only form of data available.
Age information enables targeted information dispersal and better personalization, including ensuring privacy and security, thereby enhancing the experiences supported by the speech technology applications. 
Arguably, child centric applications have more to benefit by using paralinguistic information in enabling novel approaches in safeguarding children and enforcing age appropriate content.

Most of the past research in speaker age estimation is based on adult speech.
Earlier research involved training classifiers on statistical functionals of speech descriptors such as loudness, pitch, jitter, shimmer, mel-frequency cepstral coefficients (MFCC) \cite{schuller2010interspeech,Schuller2013ParalinguisticsinSpeechand}.
Gaussian mixture models (GMM) based systems trained on MFCCs have been a popular choice for speaker age prediction \cite{li2013automatic}.
Maximum a-posteriori (MAP) adaptation, discriminative training using maximum mutual information (MMI) have been shown to be successful additions to GMMs \cite{li2013automatic,kockmann2010brno}.
\cite{kockmann2010brno} proposed joint factor analysis (JFA) with a GMM back-end for age classification.
Later, i-vector with total variability modeling trained on MFCC features significantly advanced the performance of age regression achieving 7.6 years of mean absolute error (MAE) \cite{bahari2014speaker}.
Supervised i-vectors further improved the performance by decreasing the MAE by a relative 2.4\% \cite{shivakumar2014simplified}.
Within class covariance normalization (WCCN) was found to be useful both in the case of i-vector and supervised i-vector \cite{grzybowska2016speaker,shivakumar2014simplified}.
Cosine distance scoring is typically used for classification with i-vectors and support vector regression in case of age regression task.
\cite{fedorova2015exploring} reported improvements using i-vectors by adopting shallow artificial neural networks as backend for regression.

More recently, deep neural networks have been employed for speaker age estimation.
In \cite{sadjadi2016speaker}, the hybrid acoustic DNN-HMM from an automatic speech recognition (ASR) system is used to extract phonetically-aware senone posterior i-vector, instead of the typical GMM-Universal background model (UBM).
In \cite{mallouh2018new}, bottleneck features are extracted from a hybrid DNN-HMM phone recognition system and subsequently used to train the i-vector, yielding better performance.
End-to-end deep neural network architectures have also been explored for age estimation \cite{ghahremani2018end,zazo2018age,qawaqneh2017dnn}.
One such system, popularly termed as x-vectors, comprises several layers of time delay neural network followed by a pooling layer that computes mean and standard deviation over time.
The statistics are concatenated and propagated through several feed forward layers to finally output softmax distribution over predefined, binned, age categories.
With large amounts of training data or data augmentation, x-vectors have been shown to generally outperform i-vectors for age estimation  \cite{ghahremani2018end}.
Recurrent and convolutional neural network architectures have also been explored \cite{zazo2018age,sanchez2019convolutional}.

Although, there has been interest in automatic recognition of paralinguistic information from speaker data, there has been considerably less research focused on children speech where there is significant age-dependent developmental variability \cite{lee1999acoustics,lee2014developmental}.
Most of the work involving children treat them as a broad sub-population group and perform classification across broad age groups such as children, youth, adult and senior adults \cite{li2013automatic,schuller2010interspeech,kockmann2010brno,grzybowska2016speaker,qawaqneh2017dnn}.
\cite{bocklet2008age} proposed GMM supervectors (GMM-UBM) and support vector machines (SVM) for classification and regression of children's age.
\cite{mirhassani2014age} proposed fuzzy based strategy to aggregate the output of multiple classifiers each trained using MFCC features pertaining to vowels.
Extreme learning machine and SVM was used for classification among children of 6 age classes (7 to 12 years).
In \cite{safavi2014identification} and \cite{safavi2018automatic}, children ranging from age 4 to 14 years are categorized into 3 groups based on their age and classification is performed demonstrating the performance advantage of the i-vector system trained on MFCC features and linear discriminant analysis (LDA) against GMM-UBM, GMM-SVM systems. 
In \cite{grzybowska2016speaker}, age regression is performed among the children sub-population, however the mean absolute error and the correlation in case of children is found to be poor.
\cite{sarma2020children} performed age classification task among children categorizing 4 to 14 years using deep neural network with TDNN-LSTM architecture trained on raw speech waveform.
The OGI Kids corpus was employed and data augmentation is performed using amplitude and speed perturbation to increase the training data for DNN.

There are additional challenges involved in handling and modeling children speech which complicate the process of automatic age estimation in children.
Collection of children speech data is relatively more expensive. 
The data scarcity of child speech resources poses additional challenges in statistical modeling.
Typical data augmentation techniques such as speech rate, pitch perturbations and inclusion of adult speech data which are effective tools in children speech recognition, may prove less helpful in the case of age estimation task.
One of the reasons is because children speech is characterized by age-dependent shifts in overall spectral content and formant frequencies \cite{potamianos2003robust}.
The inclusion of adult speech data is less likely to aid in the performance improvement of children age estimation because of the wide mismatch of spectral parameters between children and adult speech.
Moreover, human perceptual evaluations indicate that speech rate influences speaker age estimation \cite{skoog2015can}.
Faster speech rate is associated with lower age estimates and vice-versa \cite{skoog2015can}.
Error in age estimation is also linked to misclassification of gender \cite{assmann2015links}. For example, 
perception of gender as male portrayed tendencies toward lower age estimates \cite{assmann2015links}.
These observations make data augmentation techniques such as speech rate and pitch perturbations unsuitable for the task of automatic age estimation.

From a speech modeling perspective, child speech is relatively more complex with high signal variability due to the developmental changes along various aspects including structural (e.g., vocal tract anatomy), motoric (e.g., speech related movements), cognitive (e.g., linguistic knowledge) and social (e.g., affect expressions) \cite{lee1999acoustics,lee2014developmental}.
High within-speaker variability is observed across all children ages through adulthood \cite{lee1999acoustics}.
Substantial variation in growth rates of children reflects in substantial variation of vocal tract structure for children of same age and for a specific child at different ages \cite{vorperian2009anatomic} which further complicates modeling of children age from speech.
High inter-speaker variability observed across age groups \cite{gerosa2009review,lee1999acoustics} poses further difficulties in estimating efficient within-age class boundaries.
It is well documented in the literature that children speech recognition is significantly less accurate \cite{potamianos2003robust,gerosa2009review,shivakumar2020transfer,shivakumar2014improving,shivakumar2021end} underscoring the modeling difficulties associated with children speech.

From a psycho-acoustics perspective, the perception of children's age is particularly distinct.
Humans tend to incorporate assumptions about a child speaker's gender in estimating child's age \cite{barreda2018modeling,assmann2015links,assmann2013perception}.
In general speaker gender inference is relatively poor in case of children compared to adult speakers since speech of both female and male children is characterized with higher F0 values \cite{lee1999acoustics} which potentially manifests as errors in age estimation.
Humans were found to persistently underestimate age for older girls \cite{assmann2013perception}.
These trends pose further challenges in children speaker age estimation.

Our work is motivated from the investigations of variations of temporal parameters in children across age categories \cite{lee1999acoustics}.
\cite{lee1999acoustics} found phoneme durations to be associated with speech development in children.
In this study, we propose features derived from phone durations for the task of age estimation from children speech.
Manual transcriptions are forced-aligned with speech data to obtain phone duration distributions and are subsequently used to train regression models.
Our work remains one of the very few works in children speech domain to employ regression for speaker age estimation.
Although, there have been a few past works that incorporate durations in terms of pauses and overall length of utterances in speaker age prediction, our work is distinct in its explicit modeling of phone duration in children speech as a biomarker for speaker age estimation.
To the best of our knowledge, this work is a unique attempt at modeling speaker age information purely based on temporal variations in speech by means of phoneme duration modeling.

The rest of the paper is organized as follows: Section~\ref{sec:feats} describes the proposed phone-duration features and section~\ref{sec:model} presents the regression model.
Section~\ref{sec:data} lists the speech databases employed in our study.
The experimental setup is described in section~\ref{sec:exp_setup}.
The results are presented and discussed in section~\ref{sec:results}.
Finally, the study conclusions are provided in section~\ref{sec:conclusion} and possible future directions discussed.

\section{Phone duration features}\label{sec:feats}
Duration is a critical descriptor of speech signals.
Varying resolution of duration can convey a variety of information ranging from low level descriptors such as speaking rate and pauses in speech to more abstract information about cognitive process, emotion and conversational dynamics.
\cite{lee1999acoustics} studied the crucial role of variability of duration in children's speech.
Analysis of durations of 10 vowels and fricative portion of /S/ established significant effect of age on duration descriptors.
Younger children especially of age 5 and 6 years exhibited significantly longer mean vowel durations compared to older children, with age-dependent duration values reaching a minimum around the age 15 years.
Increased intra and inter speaker variation in duration is observed across age groups but a trend which is found to reduce with increasing age.
Both inter and intra speaker variation patterns approach adult levels for children of 13 years and above.
It was found that younger children tend to exaggerate long vowels including /IY/, /AE/, /AA/ and /ER/.
The authors also established that the effect of gender on the duration is not significant.

Few studies have established correlation between mean durations of predefined set of syllables and children speaker's true (chronological) age as well as human perceived age \cite{barreda2018modeling}.
Phrase and word durations as well as the inter-word pause durations decrease with increase in age \cite{singh2007developmental}.
Phonological studies in children link phoneme durations to developing speech articulatory and neuro-motor timing control in growing children \cite{smith1978temporal,kent1980speech,kent1976anatomical}.
\cite{lee1999acoustics} also found significant correlation between children age and sentence duration.  

Psycholinguistic studies have found that speech duration is related to cognition in children, i.e., speech patterns revealed children take longer time to express utterances with higher cognitive demand  \cite{dillon1983cognitive}.
Cognitive processes such as selection, retrieval and planning also reflect in temporal speech pause patterns found in children \cite{esposito2004children}.

Moreover, children speech is associated with increased mispronunciations, disfluencies, frequent pauses, non-vocal verbalization \cite{potamianos2003robust,potamianos1997automatic}.
Children's speech is characterized with repetitions and revisions which are reflective of language development \cite{gallagher1977revision}.
Phone duration distributions can implicitly encode such speech characteristics found in children and has the ability to capture several para-linguistic patterns including as reflected in speaking rate and stress markers.

\begin{figure}[t]
\centering
\includegraphics[width=0.9\columnwidth]{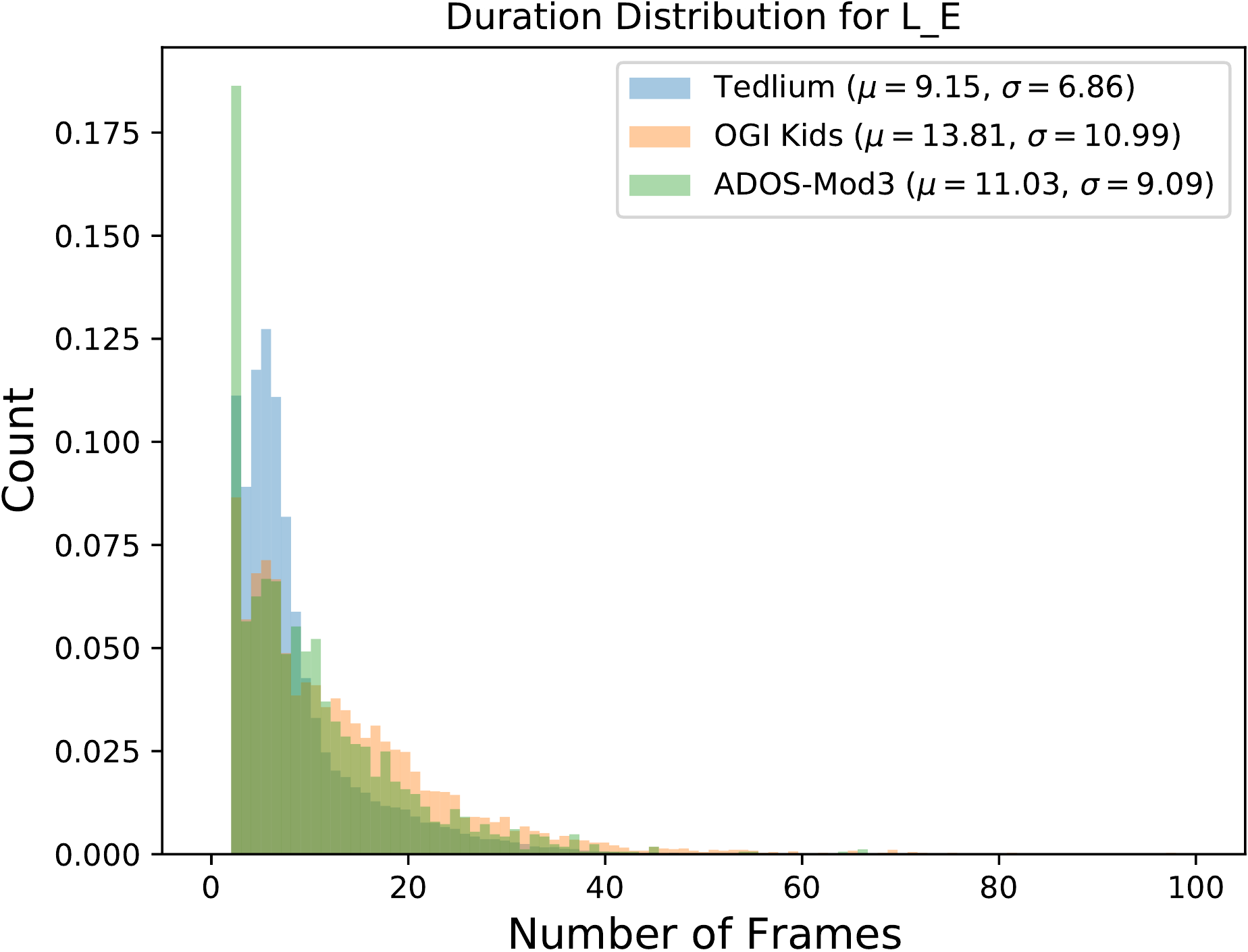}
\caption{Phone duration distribution for phoneme /L/ (end position) - Adult (TEDLIUM) vs. Children (OGI Kids \& ADOS-Mod3)}\label{fig:dist}
\end{figure}

\begin{figure}[t]
\centering
\begin{tikzpicture}
\node[inner sep=0pt] at (0,0) {\includegraphics[width=0.7\columnwidth,height=0.7\columnwidth]{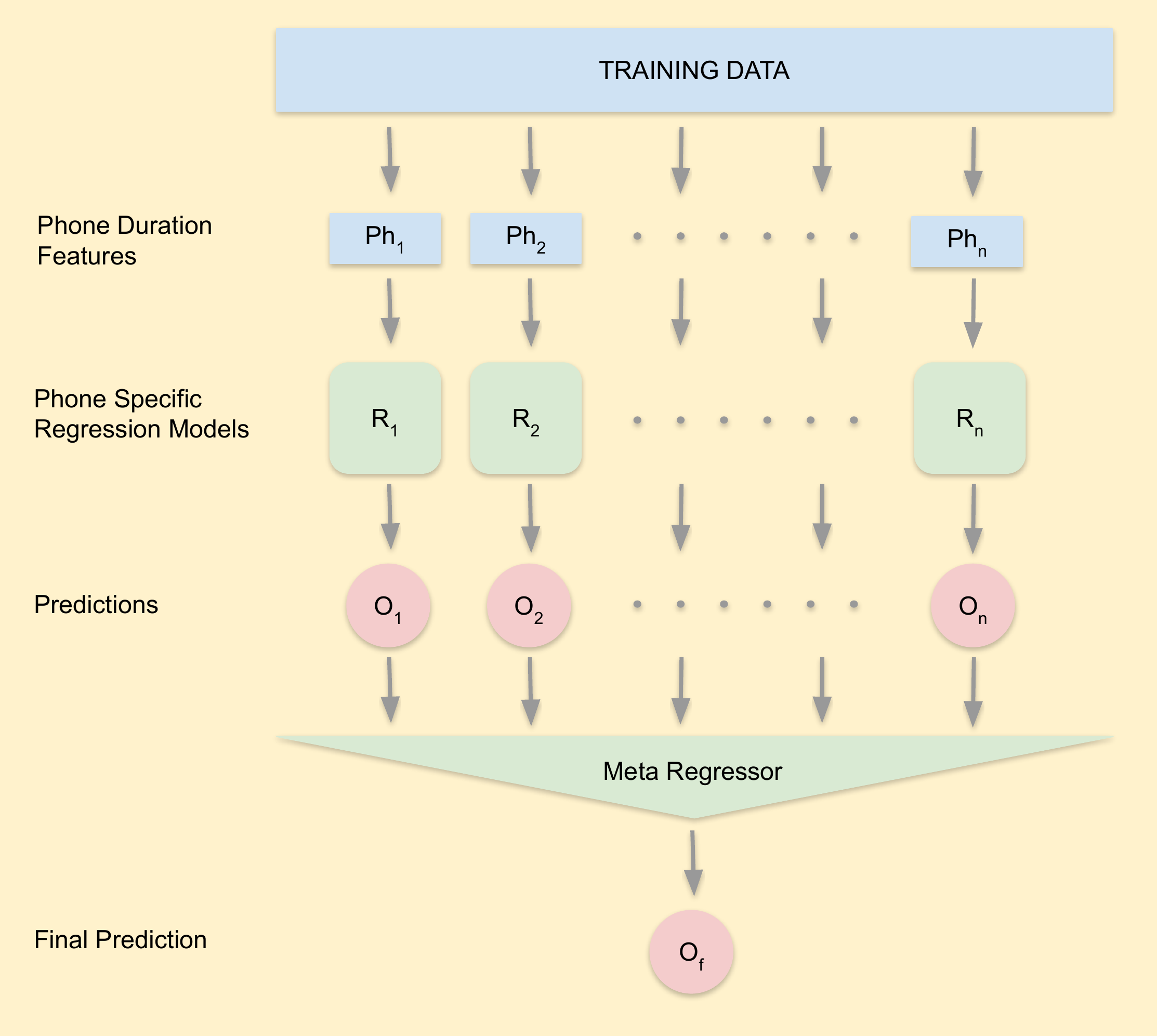}};
\draw [white, rounded corners=.3cm, line width=.2cm]
    (current bounding box.north west) --
		(current bounding box.north east) --
		(current bounding box.south east) --
		(current bounding box.south west) -- cycle
		;
\end{tikzpicture}
\caption{Proposed Age Regression Model Architecture}\label{fig:model}
\end{figure}

\subsection{Proposed Phone-Duration Features}
Motivated and supported from the findings in prior literature, in this work, we propose features explicitly engineered to model phone duration distribution in children speech to determine a speaker's age.
First, the speech data is forced-aligned with the manual transcriptions.
Later, the temporal occupancy distribution for the following set of phones are computed:
\begin{itemize}
\item Position dependent phones: to capture temporal patterns of phones depending on their position in the word (beginning, intermediate or end) or in isolation.
\item Position independent phones: obtained from aggregated statistics of position independent phones.
\item Lexical stress marked phones: vowels carrying either no stress, primary stress or secondary stress.
\item Silence phones: to model the pauses and speaking traits.
\item Special phones such as spoken noise: to model and capture hesitancy, disfluencies and filled pauses.
\item Global distributions: set of all non-silence phones, consonants and vowels.
\end{itemize}
Finally, statistical functionals are computed from the duration distributions for each phone, i.e., eight distribution descriptors namely \emph{mean, variance, minimum, maximum, skewness, kurtosis, entropy} and \emph{mean absolute deviation}.

\begin{figure*}[t]
\centering
\begin{subfigure}{0.245\linewidth}
\includegraphics[width=\linewidth]{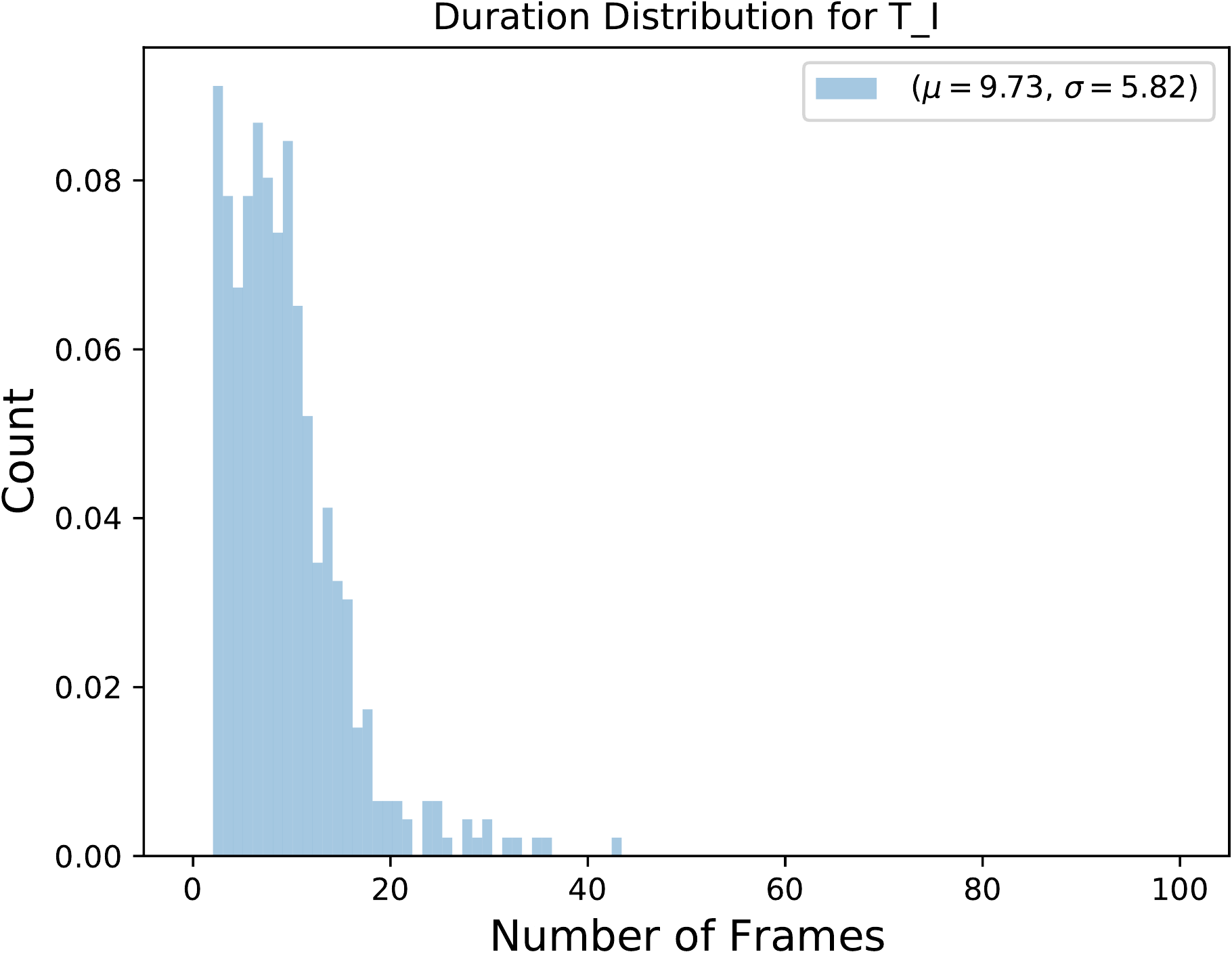}
\caption{KinderGarten}
\end{subfigure}
\bigskip
\begin{subfigure}{0.245\linewidth}
\includegraphics[width=\linewidth]{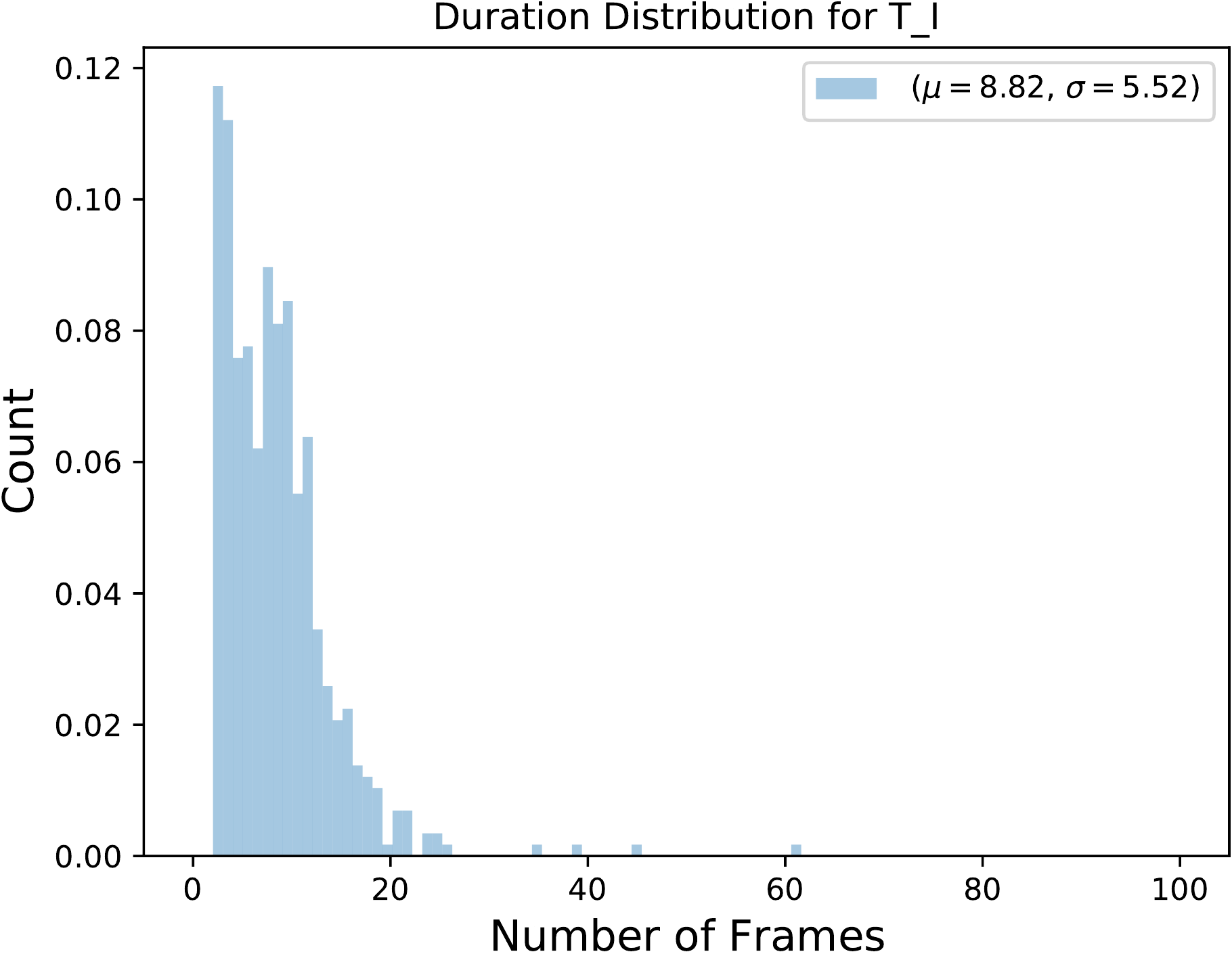}
\caption{$1^{st}$ Grade}
\end{subfigure}
\begin{subfigure}{0.245\linewidth}
\includegraphics[width=\linewidth]{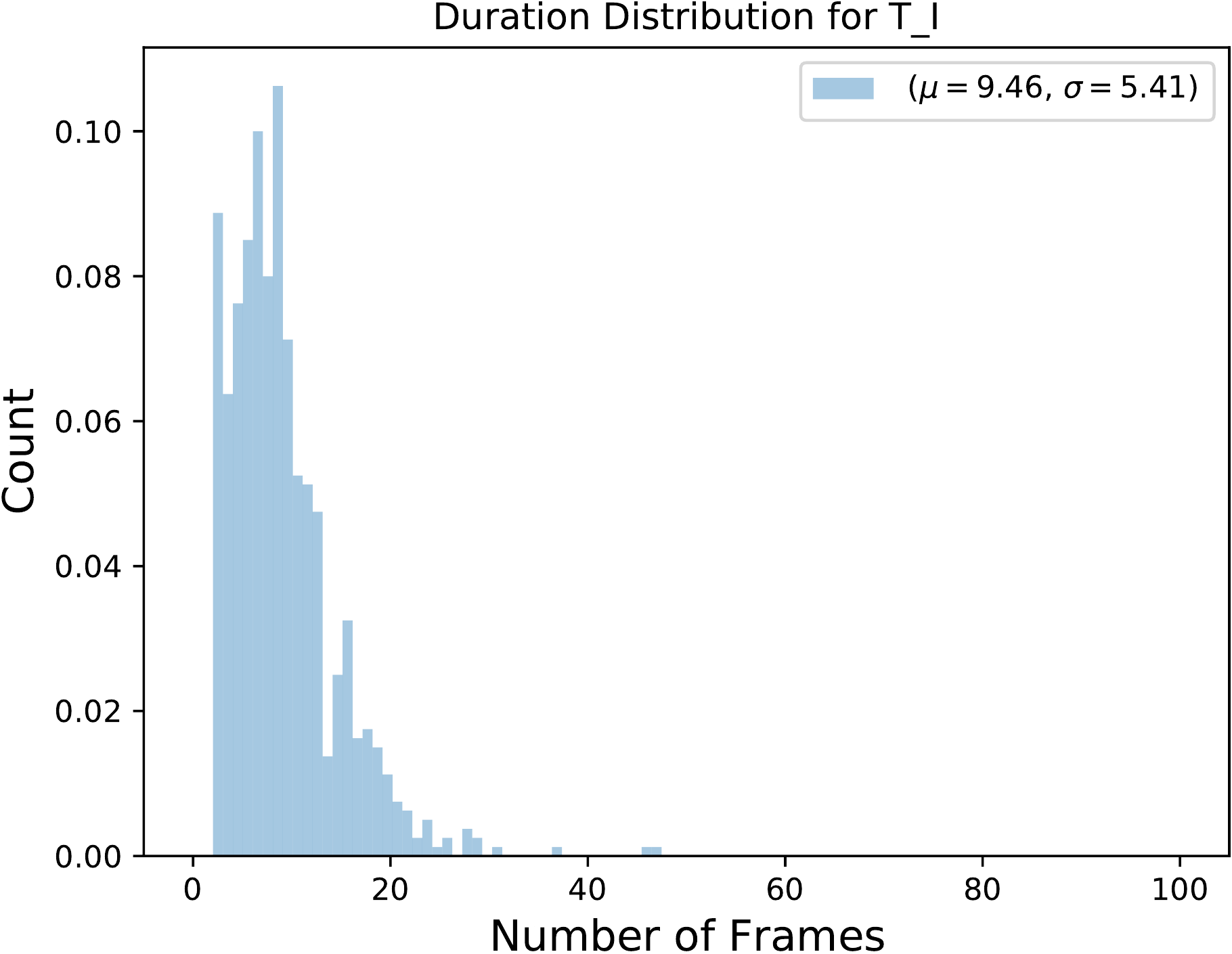}
\caption{$2^{nd}$ Grade}
\end{subfigure}
\begin{subfigure}{0.245\linewidth}
\includegraphics[width=\linewidth]{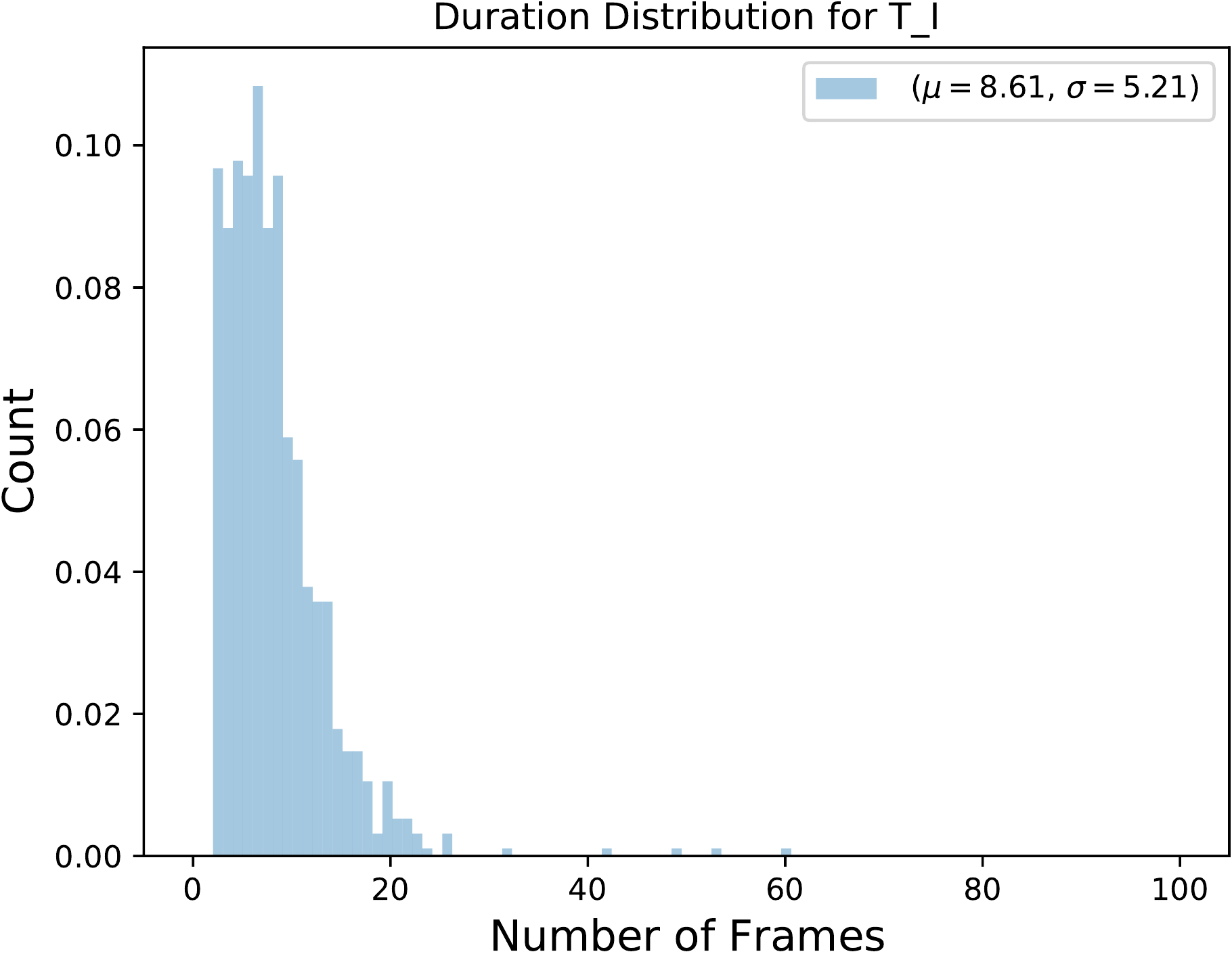}
\caption{$3^{rd}$ Grade}
\end{subfigure}
\bigskip
\begin{subfigure}{0.245\linewidth}
\includegraphics[width=\linewidth]{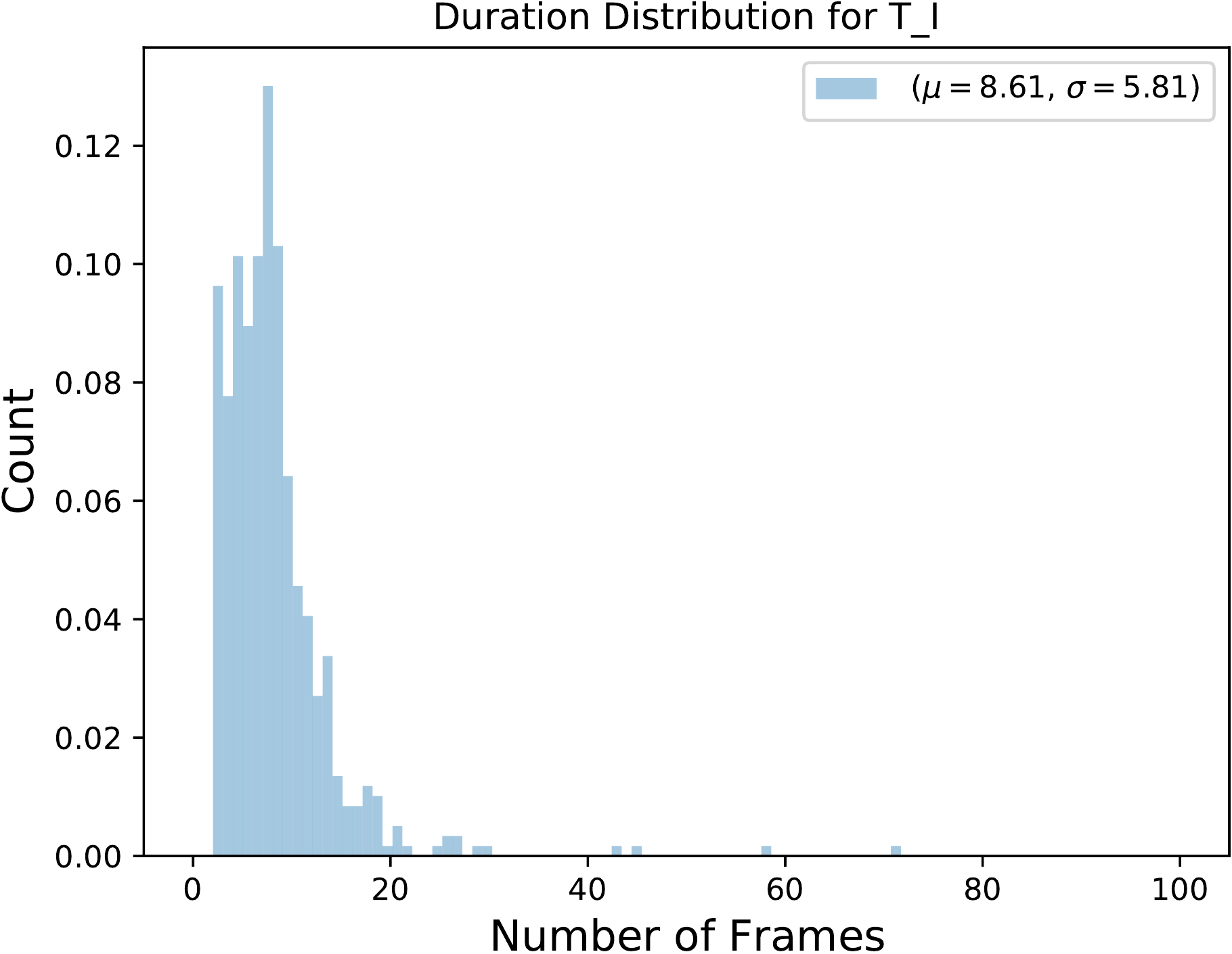}
\caption{$4^{th}$ Grade}
\end{subfigure}
\begin{subfigure}{0.245\linewidth}
\includegraphics[width=\linewidth]{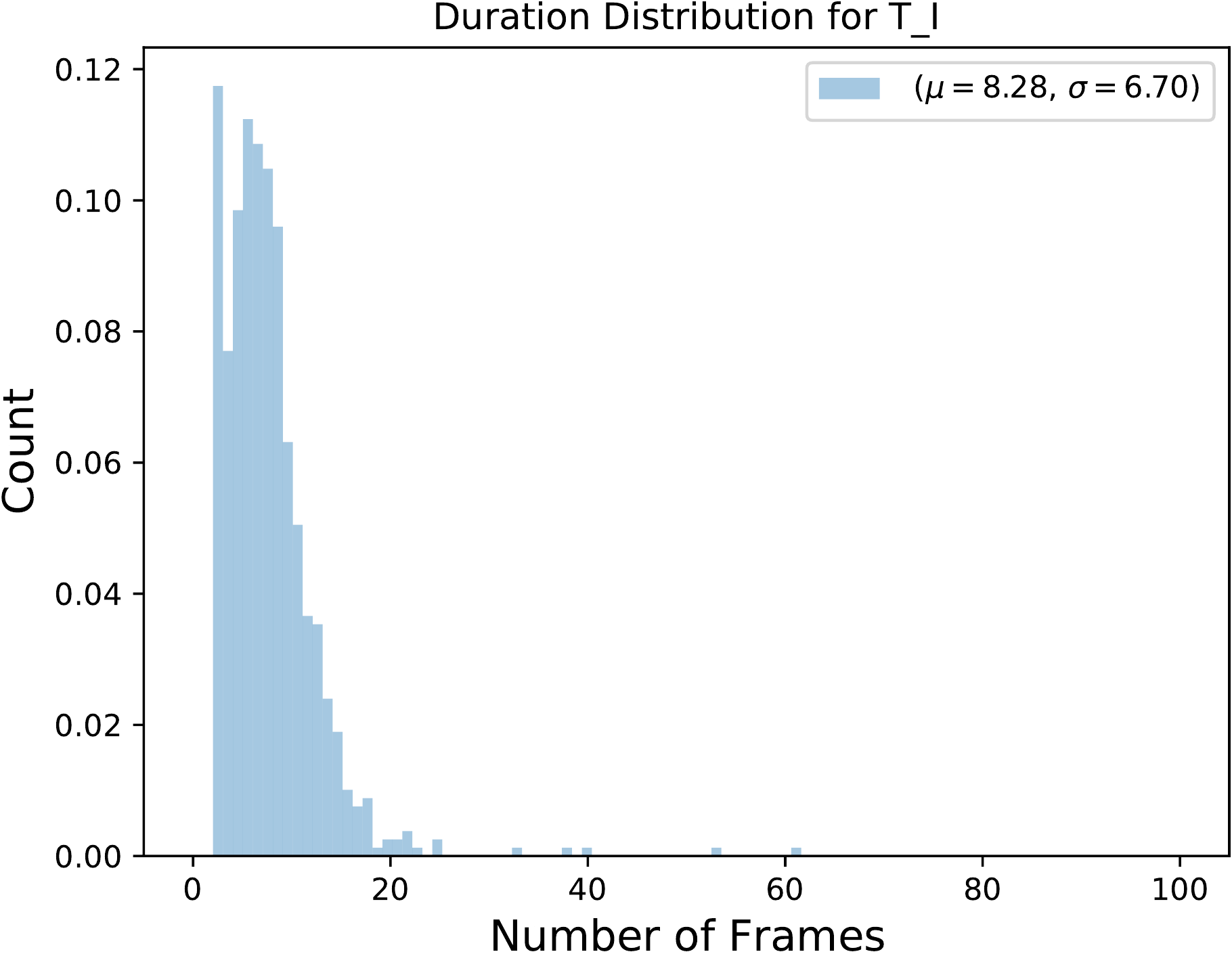}
\caption{$5^{th}$ Grade}
\end{subfigure}
\begin{subfigure}{0.245\linewidth}
\includegraphics[width=\linewidth]{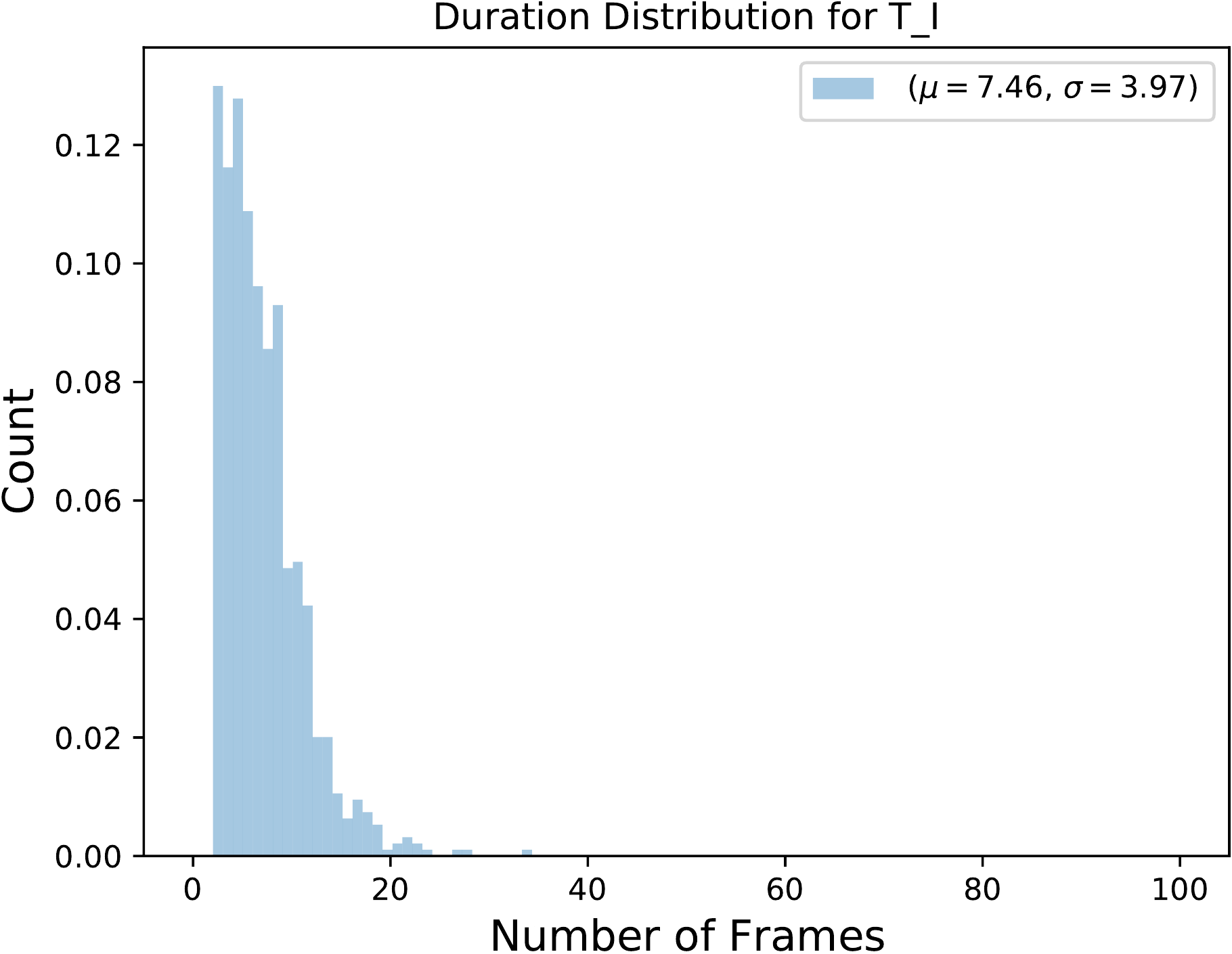}
\caption{$6^{th}$ Grade}
\end{subfigure}
\begin{subfigure}{0.245\linewidth}
\includegraphics[width=\linewidth]{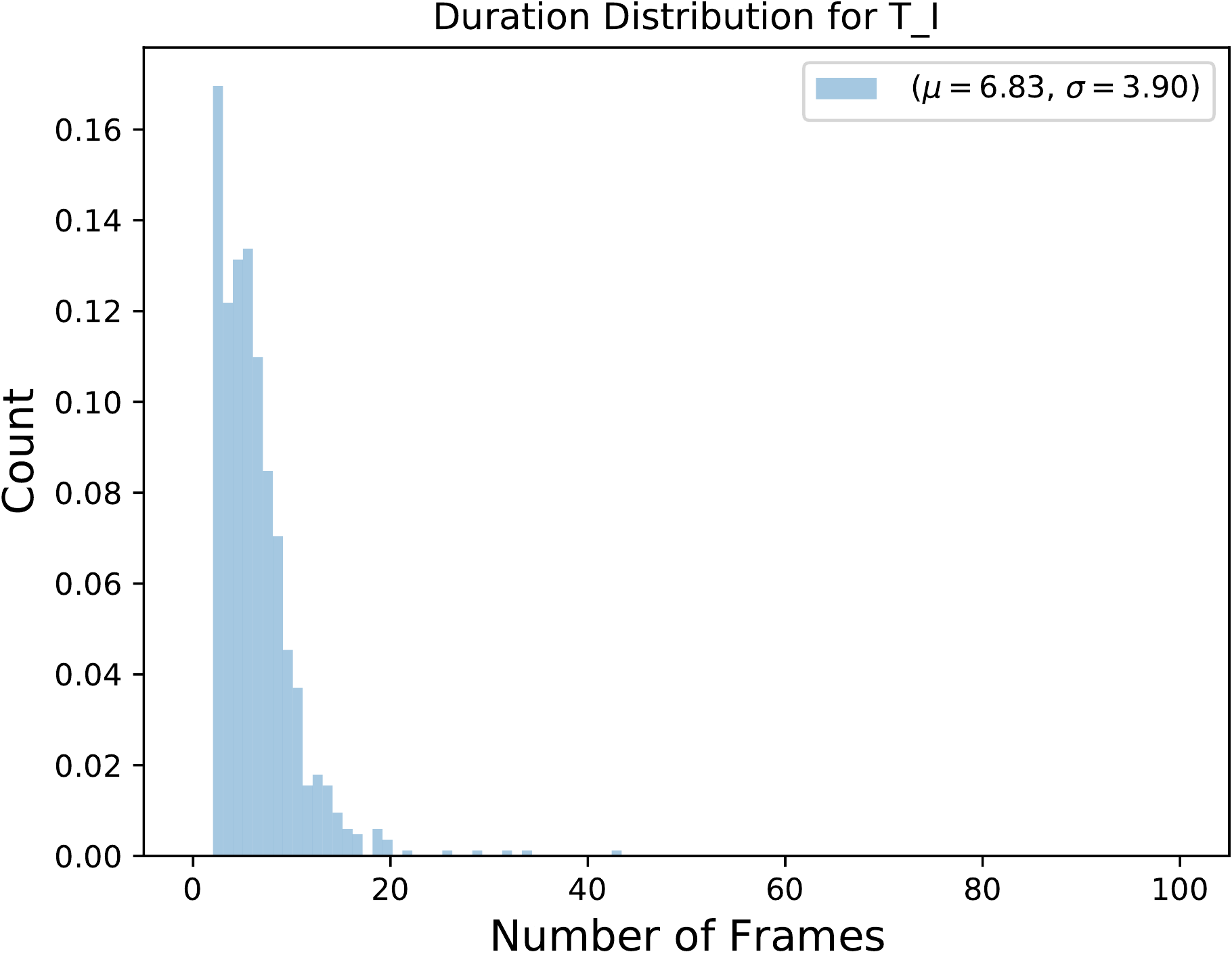}
\caption{$7^{th}$ Grade}
\end{subfigure}
\begin{subfigure}{0.245\linewidth}
\includegraphics[width=\linewidth]{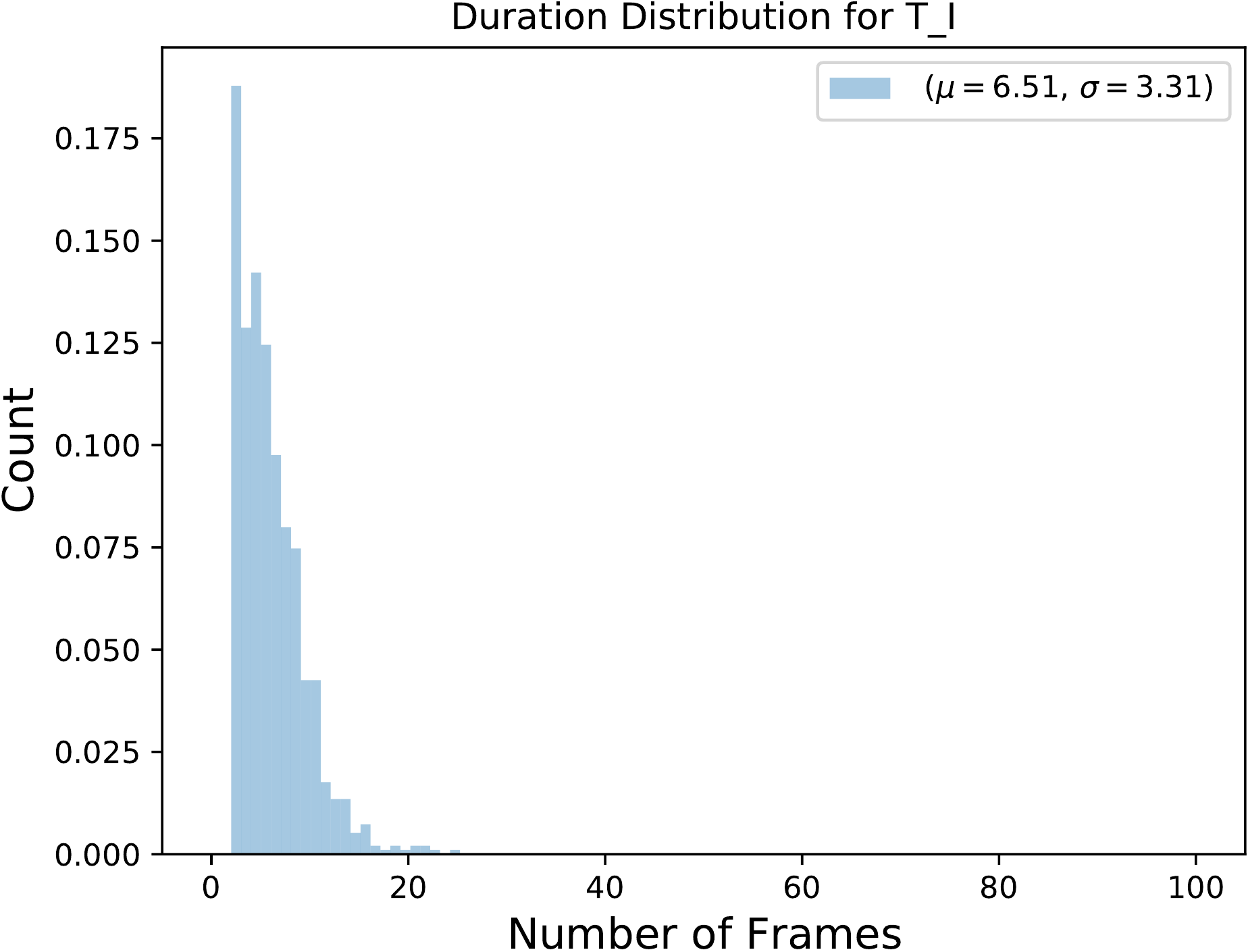}
\caption{$8^{th}$ Grade}
\end{subfigure}
\begin{subfigure}{0.25\linewidth}
\includegraphics[width=\linewidth]{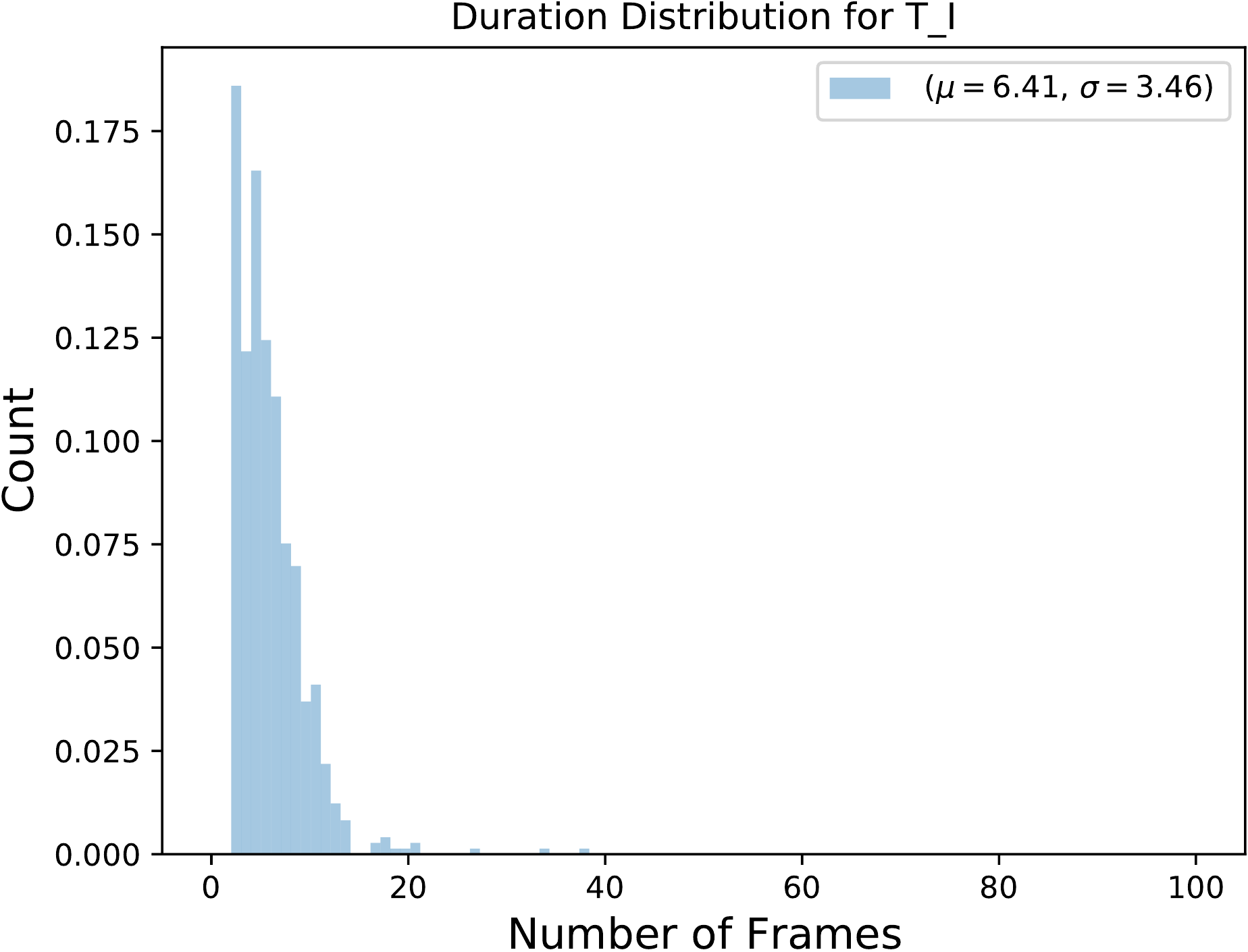}
\caption{$9^{th}$ Grade}
\end{subfigure}
\begin{subfigure}{0.25\linewidth}
\includegraphics[width=\linewidth]{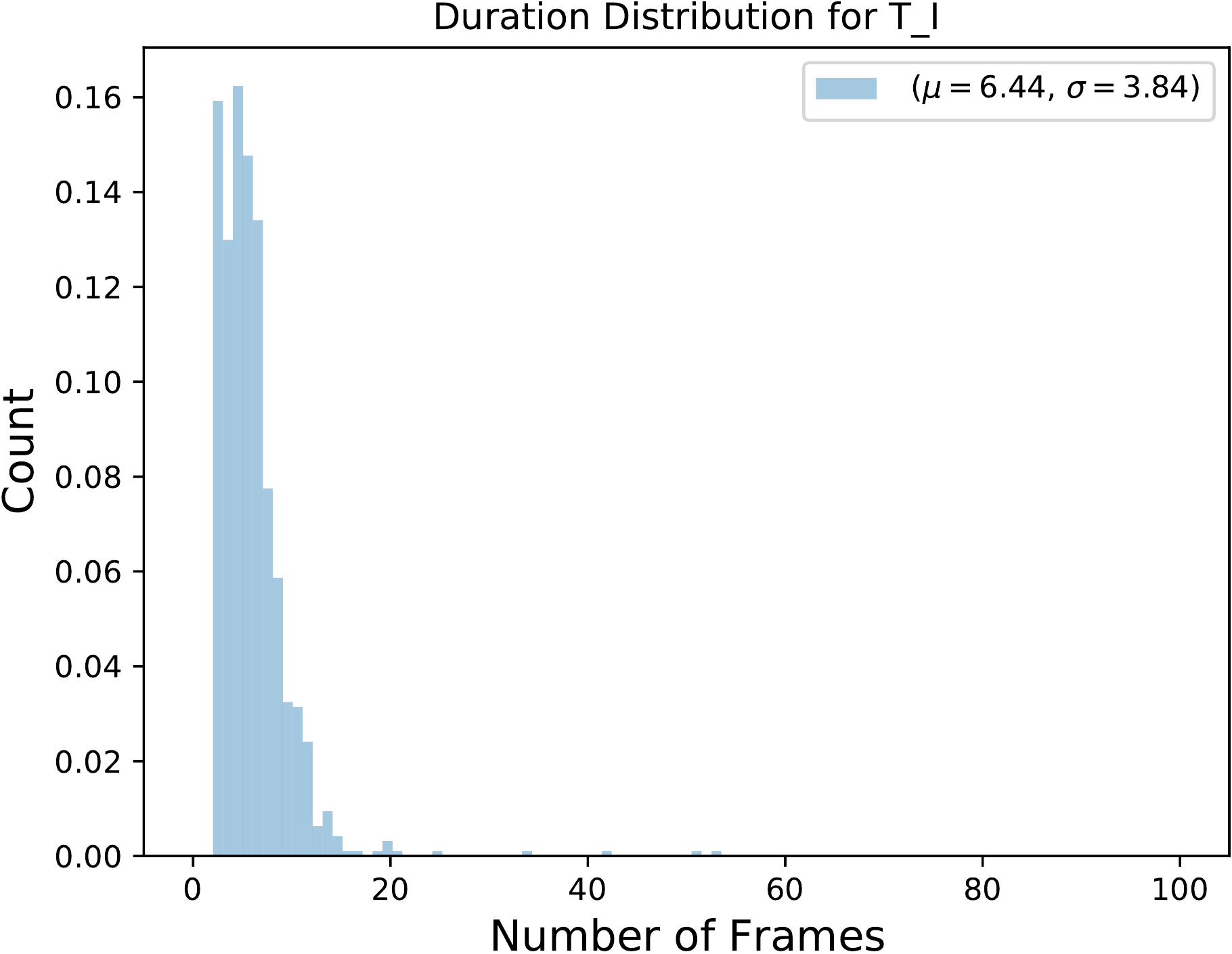}
\caption{$10^{th}$ Grade}
\end{subfigure}
\caption{Phone Duration Distribution for phone T\_I over different ages}\label{fig:dist_age}
\end{figure*}

Figure~\ref{fig:dist} shows an example of duration distribution for the position dependent phoneme L\_E, where ``\_E'' indicates the occurrence of the phone /L/ at the end of the word.
The figure comprises histograms comparing phone duration distribution for 3 different corpora: one adult (Tedlium) and two children (OGI Kids and ADOS-Mod3; see section~\ref{sec:data} for descriptions).
It is evident from the figure, that adult speech is associated with significantly smaller durations compared to children speech.
Children phone durations have typically higher means and standard deviations.
Such developmental trends with phoneme durations are prevalent among children of different age categories.
Figure~\ref{fig:dist_age} presents the phone duration distributions of phoneme /T\_I/ (I represents intermediate position of phoneme in words) for each age category ranging from kindergarten to children studying in $10^{\text{th}}$ grade.
The distribution of all the phoneme durations are found to be Gaussian in nature and unimodal.

\subsection{Proposed Age Regression System}\label{sec:model}
The proposed regression model architecture is illustrated in Figure~\ref{fig:model}.
The architecture is based on stacking two layers of regressors.
An ensemble of individual estimators, handling one phoneme each, are trained independently on the eight aforementioned distribution feature descriptors to predict the speaker's age.
A final, meta regressor operates on the outputs of the individual estimators to give the final age prediction.
The final estimator is trained on the predictions of individual estimators using cross-validation.
In this work, we employ two regression models, support vector regressor and random-forest based AdaBoost regressor.
The final, meta regressor model is of the same class as the base individual estimator.
The choice of regression models are based on the following factors: (i) the amount of training data available; less data makes DNN based models unsuitable, (ii) support vector based models are popular choice among prior literature, and (iii) decision tree based model for inferring feature importance.

The proposed stacking ensemble learning has certain advantages over a single regression model.
Ensemble learning helps in achieving low bias and low variance in final predictions.
The stacked estimators also help handle high feature dimension (2912 features, i.e., 364 phonemes, 8 features each) efficiently in contrast to typical dimension reduction alternatives.
The stacked architecture for duration modeling helps in alleviating over-fitting issues since the final estimator is trained on the cross-validated predictions of the base estimators.
It enables implicit feature selection among different phonemes, since the meta classifier operates on top of outputs of base estimators pertaining to each phoneme.
It also has the added advantage to enable easier assessment of contribution of information provided by duration distribution of underlying phone towards age estimation.

\begin{figure*}[t]
\centering
\begin{subfigure}{0.32\linewidth}
\includegraphics[width=\linewidth]{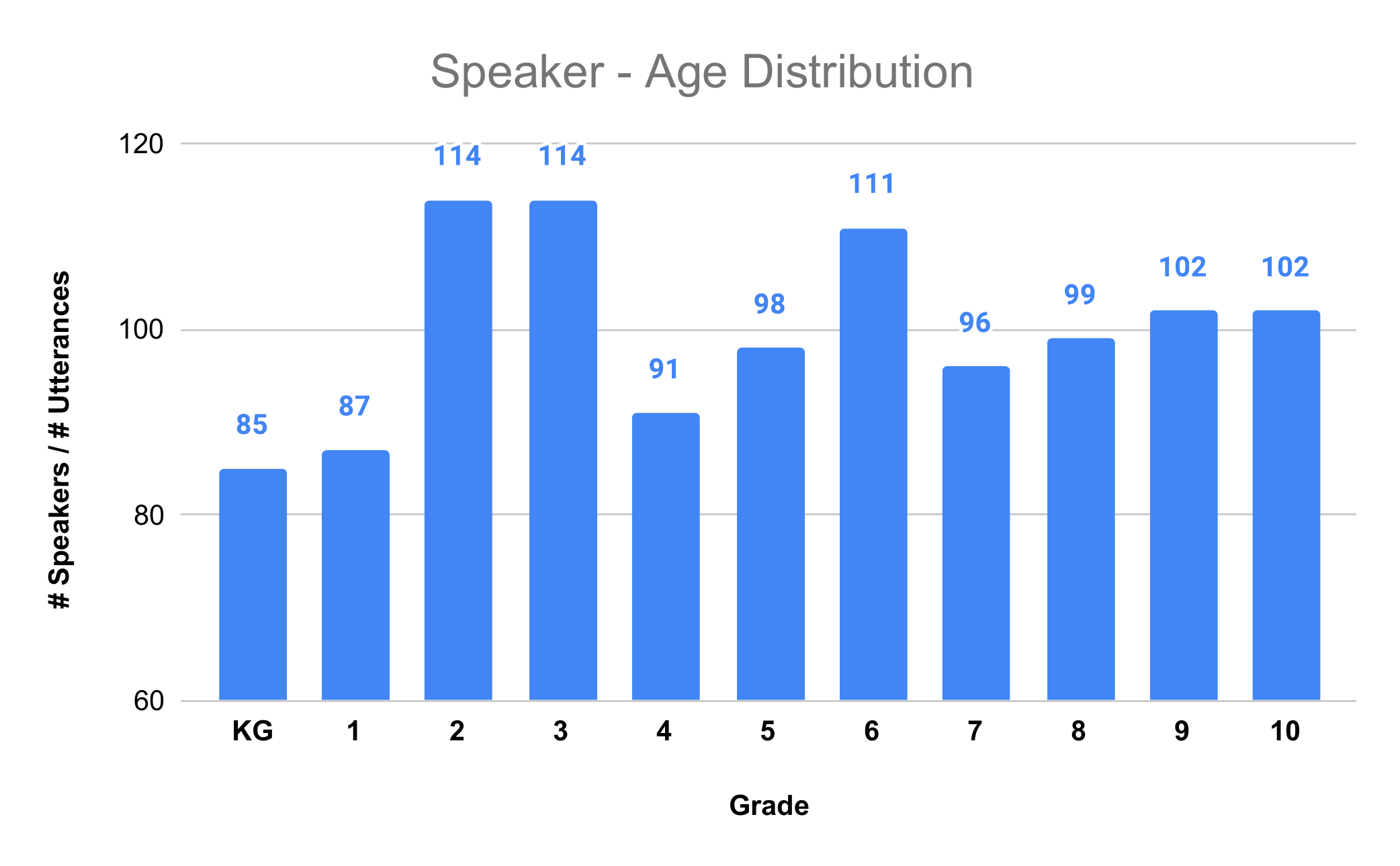}
\end{subfigure}
\begin{subfigure}{0.32\linewidth}
\includegraphics[width=\linewidth]{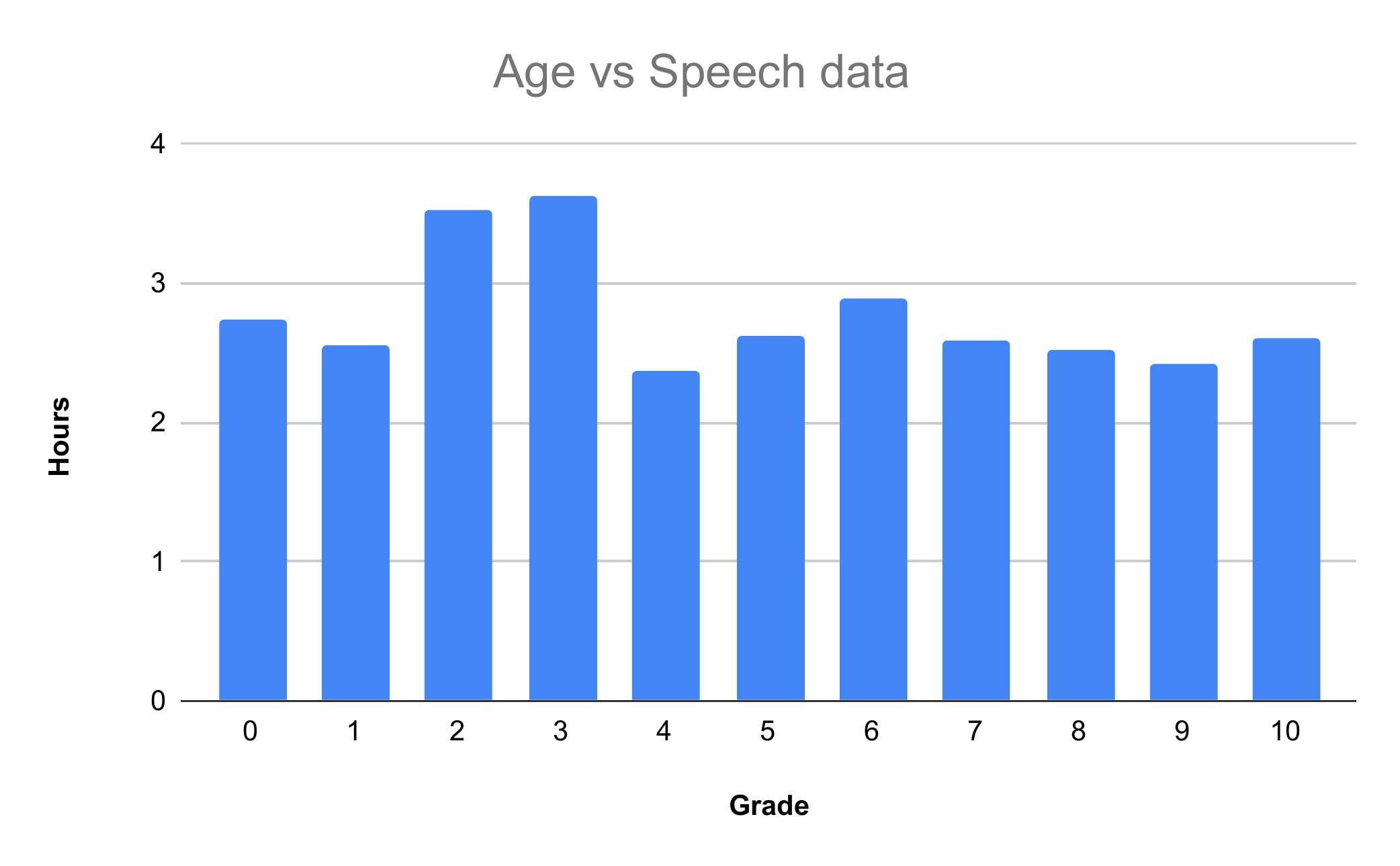}
\end{subfigure}
\begin{subfigure}{0.32\linewidth}
\includegraphics[width=\linewidth]{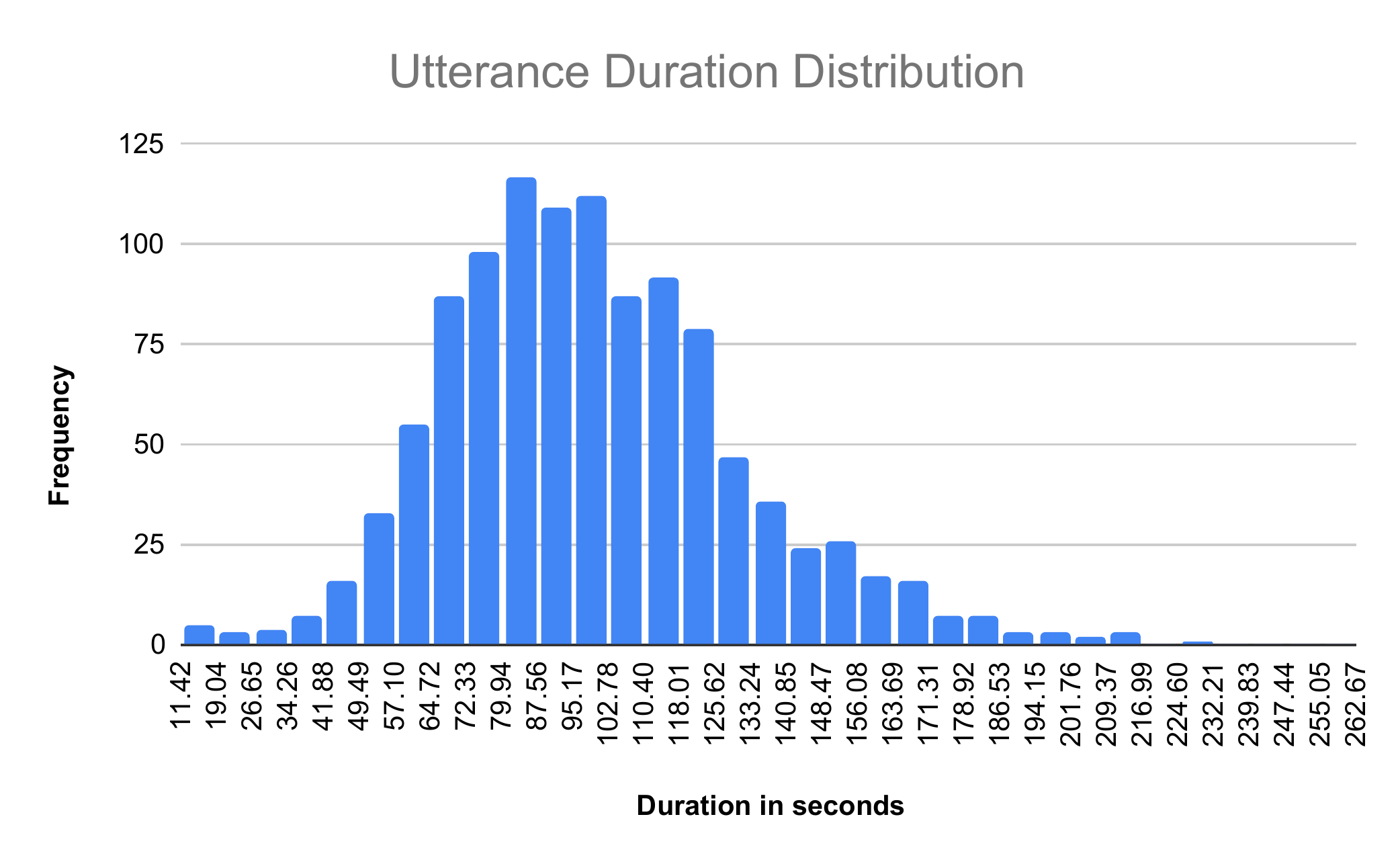}
\end{subfigure}
\caption{OGI Kids Corpus Statistics}\label{fig:ogi}
\end{figure*}

\begin{figure*}[t]
\centering
\begin{subfigure}{0.32\linewidth}
\includegraphics[width=\linewidth]{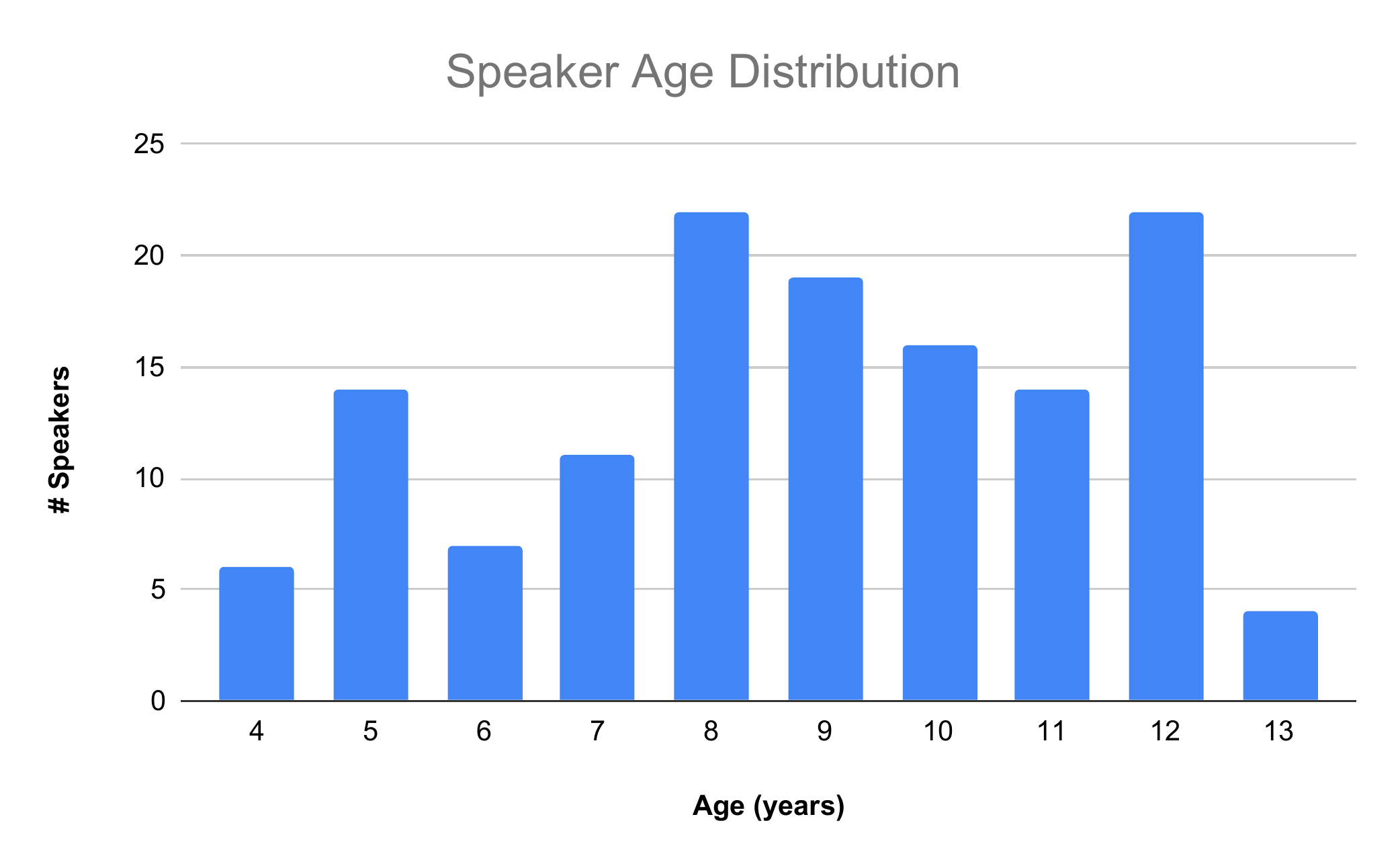}
\end{subfigure}
\begin{subfigure}{0.32\linewidth}
\includegraphics[width=\linewidth]{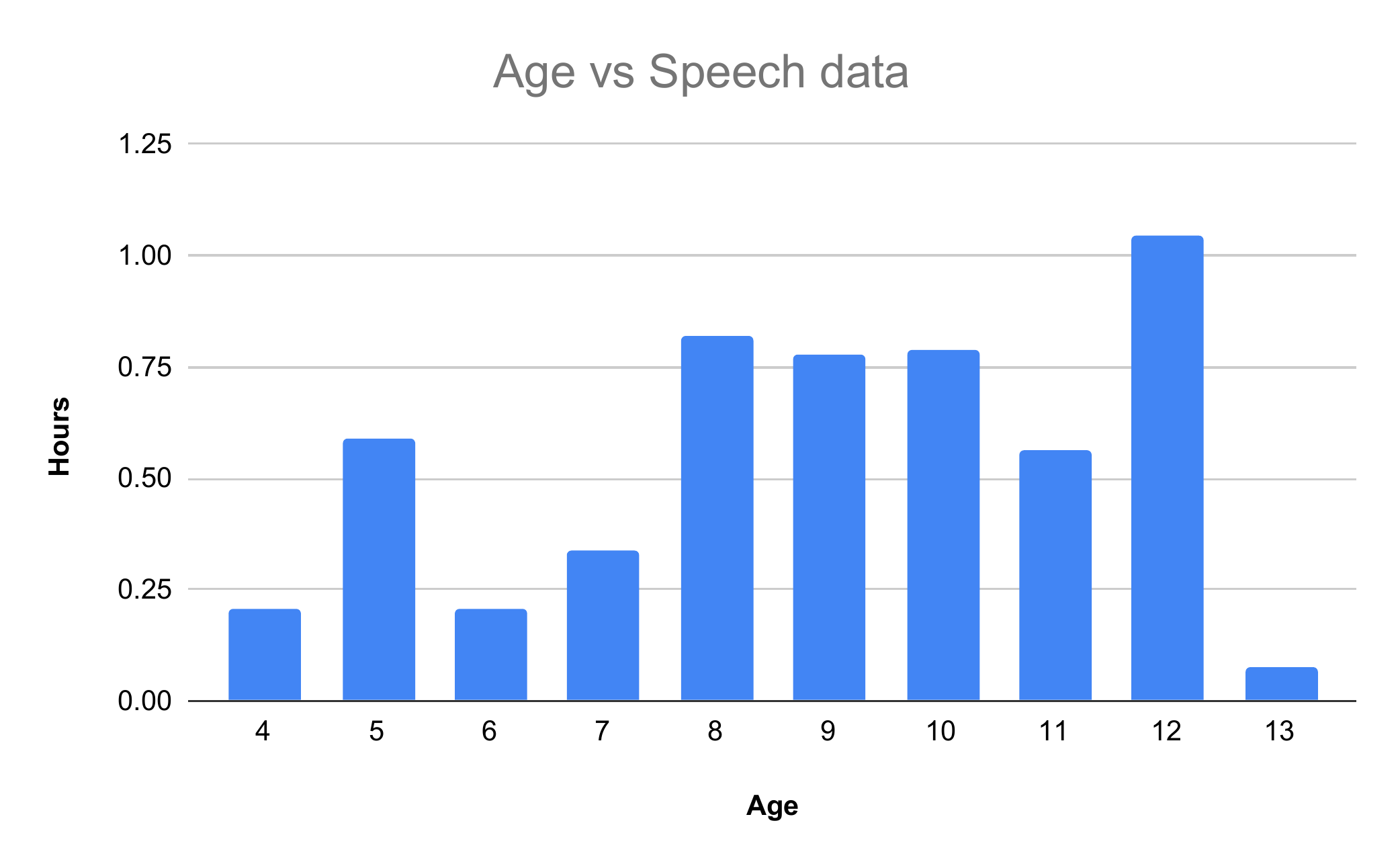}
\end{subfigure}
\begin{subfigure}{0.32\linewidth}
\includegraphics[width=\linewidth]{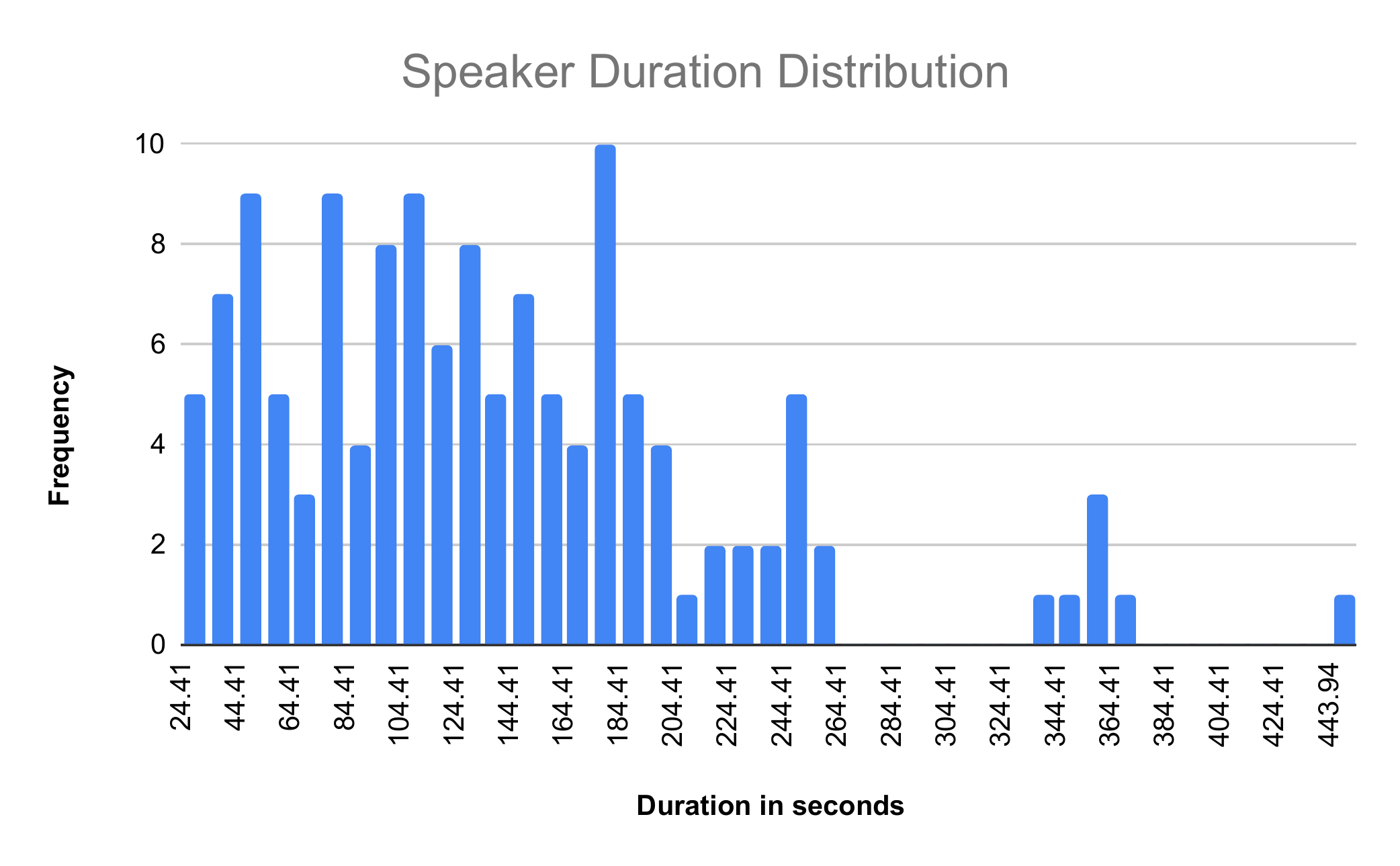}
\end{subfigure}
\caption{ADOS-Mod3 Data Statistics}\label{fig:ados}
\end{figure*}

\section{Children Speech Databases}\label{sec:data}
In this study, we perform experiments on two children speech corpora to assess the transferability and robustness to different domains and acoustic conditions.

\subsection{OGI Kids Speech Corpus}
We employ OGI Kids speech corpus \cite{shobaki2000ogi} as the primary database for our experiments due to its wide age distribution demographics among children.
We make use of the spontaneous speech subset of the corpus comprising adult interviewer asking a series of questions and eliciting a spontaneous response from children.
The corpus consists of 1100 distinct children speakers with ages ranging from children studying in kindergarten to 10th grade.
It includes a total of approximately 30.5 hours of speech.
Each speaker utters a single sentence and the mean duration per utterance is approximately 100 seconds.
The speaker age distribution, amount of speech data per age and utterance duration distribution statistics are presented in figure~\ref{fig:ogi}.

\subsection{Autism Diagnostic Observation Schedule - Module 3 (ADOS-Mod3)}
The ADOS-Mod3 corpus \cite{lord2000autism} comprises child-adult dyadic conversations involving semi-structured, standardized assessment of communication and social interactions.
The children in the corpus are diagnosed with autism spectrum disorder (ASD), attention deficit hyperactivity disorder (ADHD) and various other developmental disorders including language disorder.
The speech sessions were collected at two different locations including University of Michigan Autism and Communication Disorder Center and Cincinnati Children's Medical Center.
We make use of speech data from children only for speaker age estimation and omit adult speech data.
The corpus consists of 179 children, out of which we consider a subset of 135 children for whom we had generated good quality automatic phonetic alignments (see section~\ref{sec:exp_setup} for details regarding alignments) for the age regression analysis of this study.
The age of children range from 43 to 158 months (4 to 13 years).
The corpus contains a total of 5.4 hours of manually-transcribed speech.
Each speaker has a mean duration of approximately 144 seconds.
The speaker age distribution, amount of speech data per age and speaker duration distribution are presented in Figure~\ref{fig:ados}.

Several factors associated with this dataset add additional complexity for the task of speaker age estimation.
First, the differences in the neuro-developmental condition due to ASD, ADHD and other developmental disorders possibly reflected in the speech complicates the age estimation from speech \cite{oller2010automated}.
Second, the speech data are recorded in far-field conditions with a single distant microphone leading contributing to acoustic variability.
Third, the data analyzed are from two different locations with different room and channel characteristics adding to the complexity of speech modeling.
Finally, the corpus consists significantly less data (17\%) compared to OGI speech corpus.
The above challenges help us evaluate the robustness of the proposed phone duration model.

\section{Experimental Setup}\label{sec:exp_setup}
In this section, we provide details of our experimental setup for reproducibility purposes.
For forced-alignment we employ the KALDI speech recognition toolkit \cite{povey2011kaldi}.
The feature pipeline consists of extracting 13-dimensional mel-filter cepstral coefficients (MFCC) using a window size of 25ms and a shift of 10ms.
Linear discriminant analysis (LDA) is performed on top of the MFCC features by considering left and right context of 3 frames. Furthermore, 
Maximum likelihood linear transform is applied on top of LDA features.
Finally, feature-space maximum likelihood linear regression (fMLLR) based speaker adaptive training is used to train a Guassian mixture model - hidden markov model (GMM-HMM) based acoustic model.
The resulting acoustic model is used for forced-alignment to obtain phoneme level alignments.
Later, the statistics are accumulated for each phoneme under consideration and their 8 functional descriptors namely \emph{mean, variance, minimum, maximum, skewness, kurtosis, entropy} and \emph{mean absolute deviation} are computed.

\begin{figure*}[t]
\centering
\begin{subfigure}{0.45\linewidth}
\includegraphics[width=\linewidth]{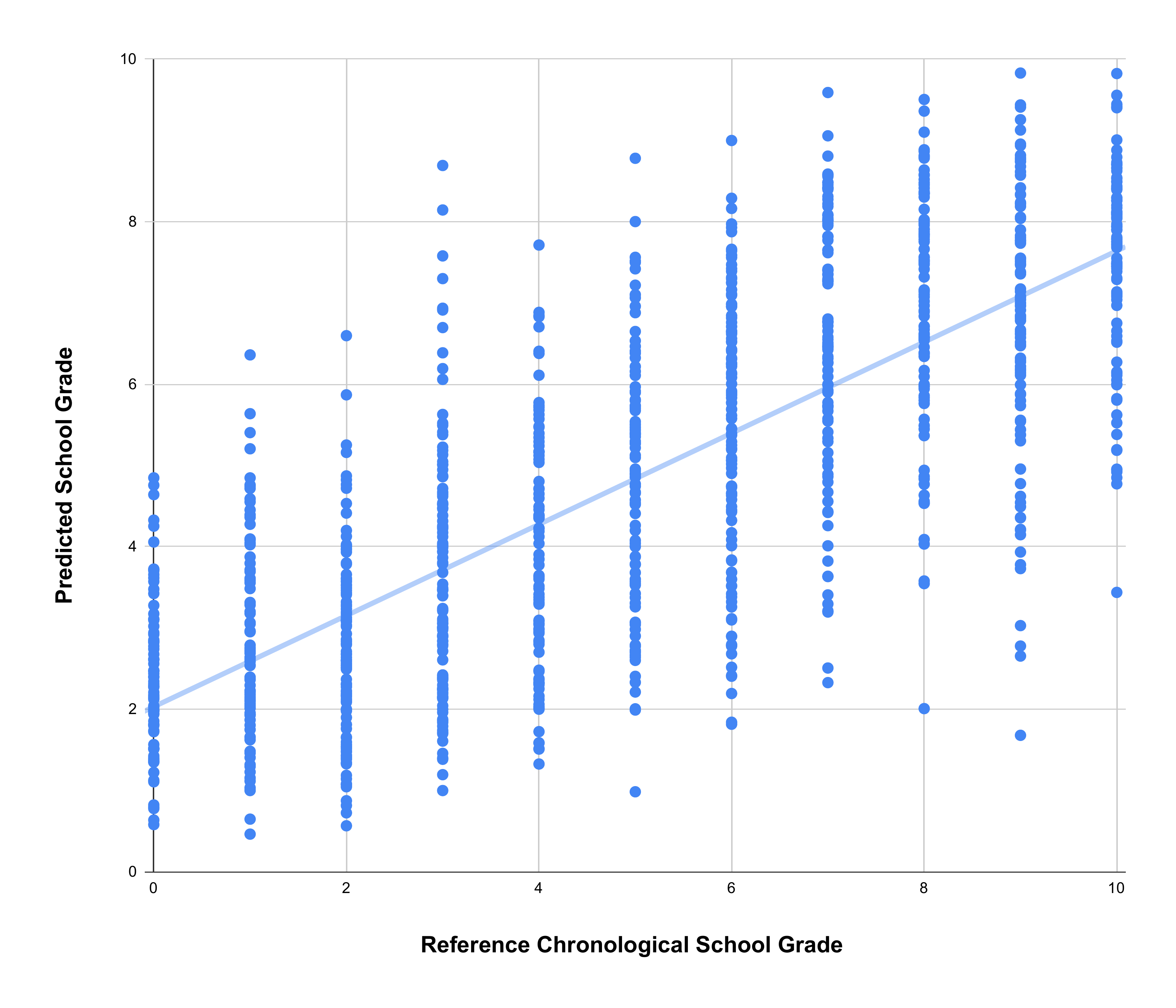}
\caption{OGI Kids}\label{fig:ogi_scatter}
\end{subfigure}\hspace{10mm}
\begin{subfigure}{0.45\linewidth}
\includegraphics[width=\linewidth]{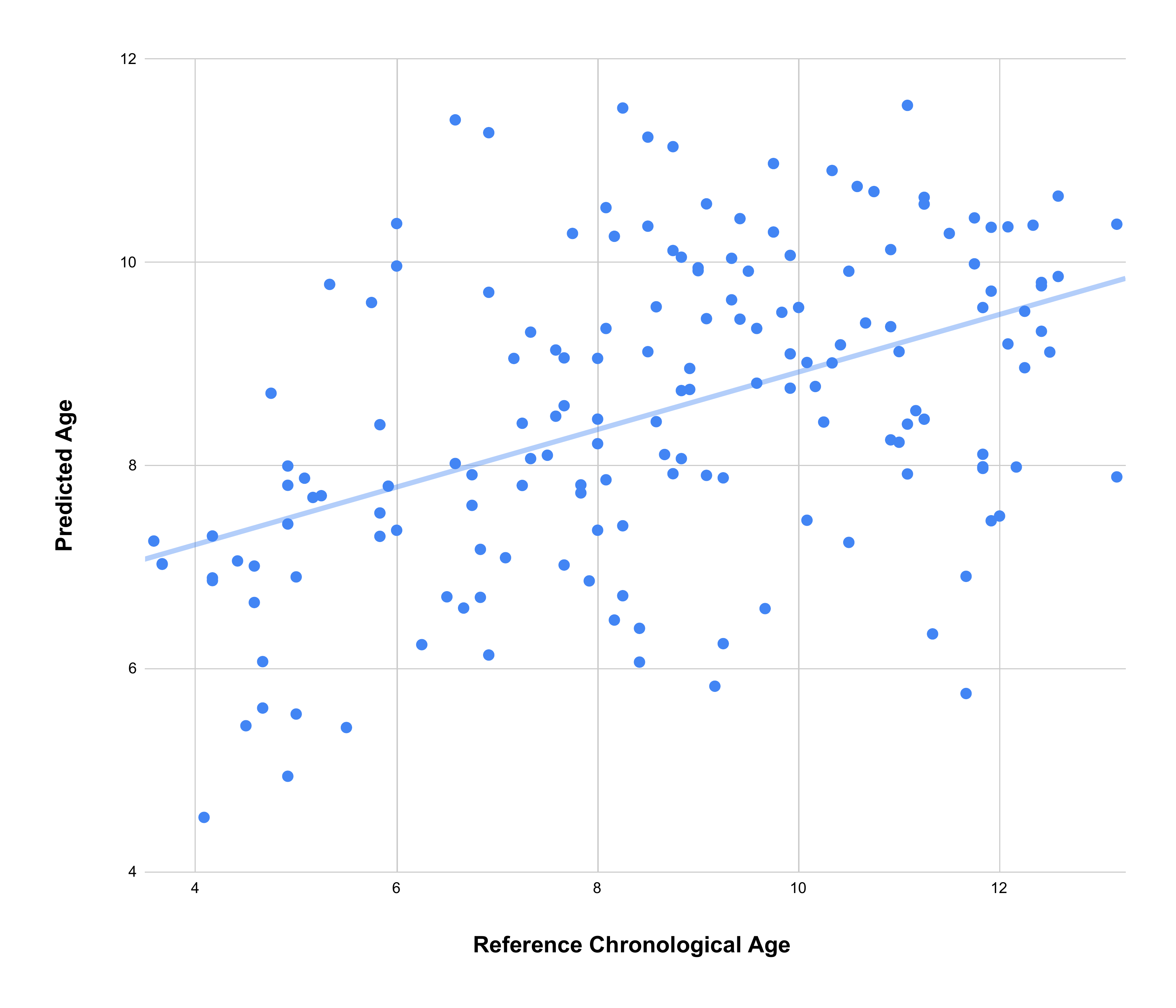}
\caption{ADOS-Mod3}\label{fig:ados_scatter}
\end{subfigure}
\caption{Age Regression Scatter Plot}\label{fig:scatter}
\end{figure*}

Two separate acoustic models are trained for forced alignments, one for the OGI Kids Corpus and the other for the ADOS-Mod3 corpus.
Since both OGI Kids and ADOS-Mod3 corpus are fairly small, we include additional speech data in favor of constructing better acoustic models and thereby obtaining better quality alignments.
The acoustic model employed for OGI Kids is trained by including 198 hours of My Science Tutor (MyST) children speech corpus \cite{ward2011my} involving children studying in grades 3, 4 and 5.
The MyST corpus comprises conversational style speech of children recorded under low noise and close talk conditions similar to OGI Kids.
The acoustic model employed for ADOS-Mod3 is trained by including all the 173 children with the addition of adult speech of clinicians conducting the Autism diagnostic assessment.
The addition of adult speech is known to yield better quality of acoustic models under low data scenarios \cite{shivakumar2020transfer}.
We do not include the MyST data since we believe the addition of adult speech data under similar recording conditions i.e., far-field, high reverberation environment in case of ADOS-Mod3 is more beneficial.
The total number of phones in the OGI Kids corpus is 364.
Whereas, the number of phones in case of ADOS-Mod3 corpus is restricted to 185 phones (excludes lexical stress markers) to better handle the smaller size of the training corpus.

For the age estimation task, we directly perform regression to predict the (reference) school grade of children (as proxy for age) in case of OGI Kids Corpus.
In case of ADOS-Mod3, the age of children were converted from months to years.
The performances are hence directly comparable between the two models.
In this work, we experiment with two regression models, i.e., support vector machine regressor (SVR) and the decision tree based random forest AdaBoost regressor.
The choice of SVR is due to its popularity and proven effectiveness for age estimation in prior works. 
Whereas, the decision tree based AdaBoost model has the benefit of providing feature importance which helps us analyze the contributions of phonemes and their discriminative power towards age estimation. 
Given the small size of the speech corpora, we perform leave-one-speaker-out (LOSO) cross-validation.
The hyper-parameter tuning of the regression models are handled implicitly through nested cross-validation.
For performance evaluation, we report mean absolute error (MAE), $\text{R}^2$ score and Pearson correlation.
In this work, we employ a simple baseline model that predicts the mean of the age of the speakers for comparison purposes.

\begin{table}[b]
\centering
\begin{tabular}{lllll}
\toprule
Database & Model & MAE & $\text{R}^2$ score & Correlation \\
\midrule
\multirow{3}{*}{OGI Kids} & Baseline & 2.69 & 0.0 & 0.0 \\
& SVR (RBF) & 1.62 & 0.58 & 0.76 \\
& AdaBoost & 1.82 & 0.48 & 0.71 \\
\midrule
\multirow{3}{*}{ADOS-Mod3} & Baseline & 2.07 & 0.0 & 0.0 \\
& SVR (RBF) & 1.79 & 0.24 & 0.49 \\
& AdaBoost & 1.74 & 0.29 & 0.54 \\
\bottomrule
\end{tabular}
\caption{Results: Children Speaker Age Estimation}\label{tab:results}
\end{table}

\section{Results}\label{sec:results}
Table~\ref{tab:results} presents the results of children speaker age estimation on OGI Kids and ADOS-Mod3 through regression.
Our proposed phone duration model achieves a mean absolute error of 1.62 and a correlation of 0.76 with SVR on OGI Kids corpus.
The results are significantly better than the baseline system based on mean age prediction.
Moreover, the correlation results are comparable to correlation between human perceived speaker age and the chronological age which is believed to be approximately 0.7 \cite{assmann2013perception}.
We observe that the SVR model outperforms the AdaBoost model.

The results with ADOS-Mod3 are slightly worse compared to OGI Kids.
This is expected due to two factors: (i) the neuro-developmental disorders can have an impact on cognitive development, and reflected in speech production differences; this in turn can lead to differences in speaker age perception and predictions from the acoustic speech signal, and (ii) the ADOS-Mod3 corpus (5.4 hours) has significantly less data compared to OGI Kids (30.5 hours).
However, the results are significantly better than the mean baseline.
In the case of ADOS-Mod3, we observe that the AdaBoost regression model outperforms the SVR.
The results on ADOS-Mod3 corpus further attests to the robustness of the proposed phone duration modeling.
Figures~\ref{fig:ogi_scatter} and~\ref{fig:ados_scatter} illustrate scatter plots of chronological age versus the predicted speaker age on the OGI Kids and ADOS-Mod3 corpora, respectively.
Note, the scatter plots for OGI Kids are quantized to the school grade of children, whereas the ADOS-Mod3 has age in terms of months.
Overall, the results suggest that the proposed phone duration features contain developmental information in children speech.
This emphasizes that speaker age in children can be robustly estimated by modeling the temporal variation via phone duration distributions.

\begin{figure}[t]
\centering
\includegraphics[width=\columnwidth,height=0.65\columnwidth]{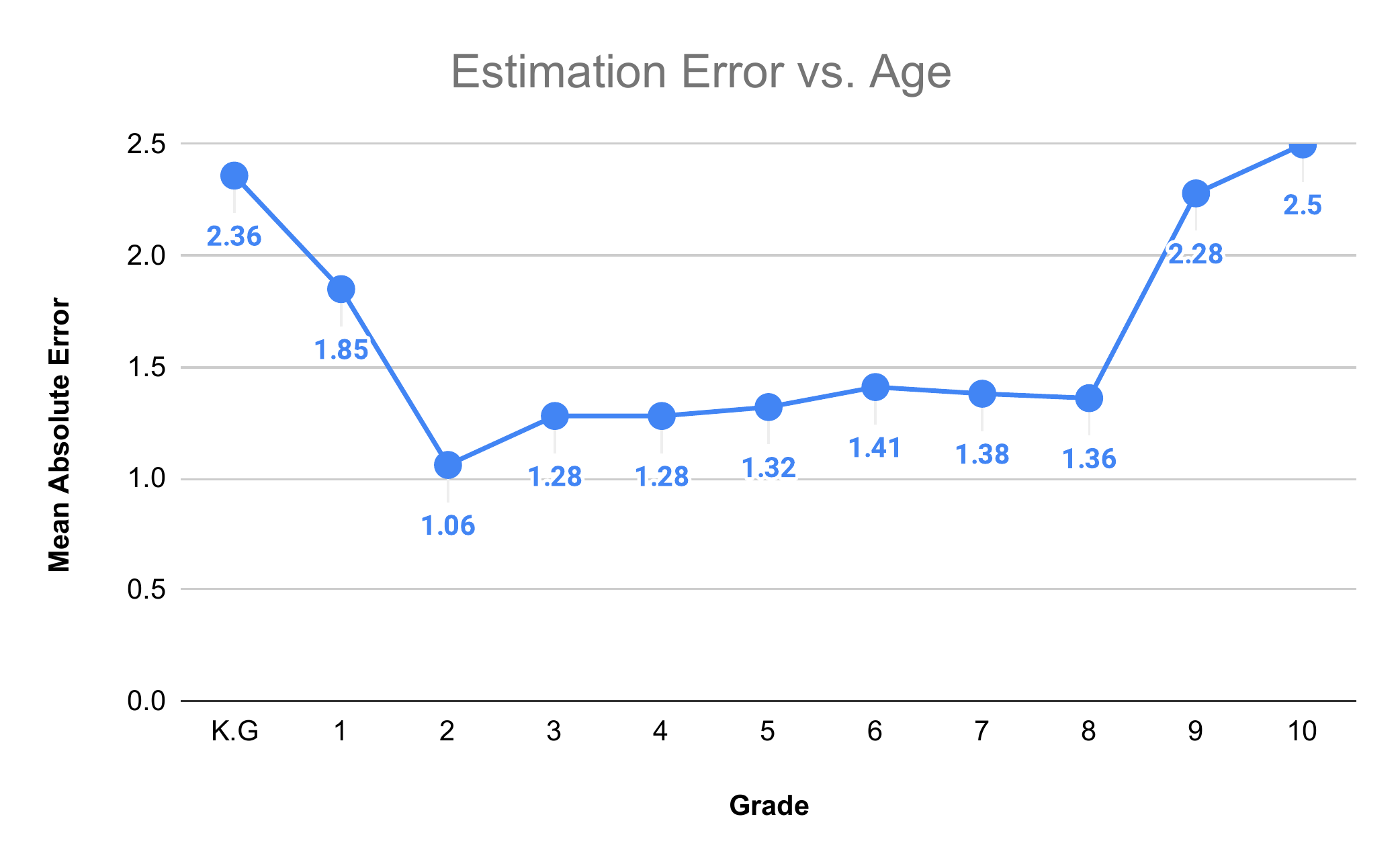}
\caption{MAE across different age categories - OGI Kids Corpus}\label{fig:mae_vs_age}
\end{figure}

Next, we perform additional analysis to assess the factor of age on the performance of the system.
Figure~\ref{fig:mae_vs_age} plots the mean absolute error in each age category of OGI kids corpus using the SVR model.
From the results, we observe that the error is low for children ages ranging from $2^{\text{nd}}$ grade to $8^{\text{th}}$ grade categories, reaching minimum for the $2^{\text{nd}}$ grade group.
However, the error increases sharply for younger (kindergarten and $1^{\text{st}}$ grade) and older children in the corpus ($9^{\text{th}}$ - $10^{\text{th}}$ grade).
One possible explanation for the observed trend is as follows: (i) in case of younger children studying in kindergarten and $1^{st}$ grade, although the inter and intra age variations are expected to be very high, the intra-age variation dominates, thereby resulting in high error, (ii) in case of children studying among $2^{nd}$ and $8^{th}$ grade, the inter-age variation dominates resulting in lower error rates, and (iii) in case of elder children ($9^{th}$ and $10^{th}$ grade) both the intra and inter age variations are significantly less which results in relatively low performance (comparable to that of adults).

\begin{figure}[t]
\centering
\includegraphics[width=\columnwidth,height=0.65\columnwidth]{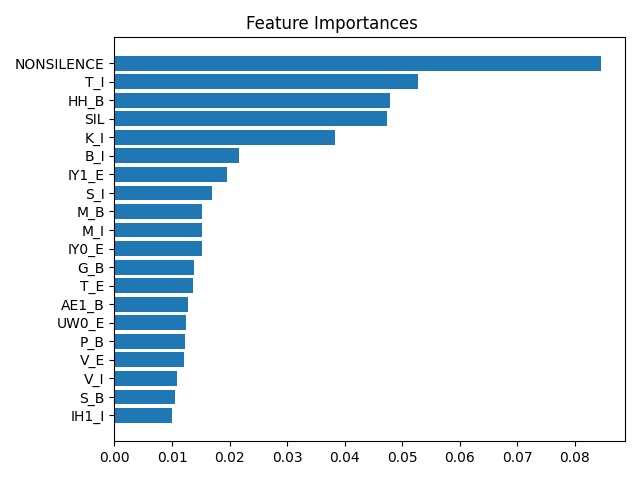}
\caption{Phoneme-wise feature importance - OGI Kids Corpus}\label{fig:feat_imp}
\end{figure}

We also perform feature importance analysis to gain insights on the contributions of each phoneme toward children speaker age estimation.
We derive impurity based feature importance that are accessible from tree-based algorithms, in our case the random forest based AdaBoost model trained on OGI Kids corpus.
The importance measures are computed based on total reduction of the optimization criterion, often referred to as the Gini importance.
We compute the feature importance only on the final, meta estimator that operates on the output of the phoneme specific base estimators.
This allows us to assess the contribution of input features on the phoneme level rather than statistical functionals.
Figure~\ref{fig:feat_imp} shows the bar plot of the top-20 most contributing phonemes computed on OGI Kids corpus.
Higher values translate to more importance. 
The following observations are made with our experimental setup:
\begin{enumerate}
\item /NONSILENCE/ (aggregated duration of {\em all} position dependent non-silent phones) comes out as the most critical contributor for speaker age estimation.
\item Position dependent phones /T\_I/, /HH\_B/ and /K\_I/ contribute critically to determining speaker age in children.
\item /SIL/ (silence) capturing inter-word pauses, speaking rate, and disfluencies also helps in determining children age.
\item The above 5 acoustic categories are more than twice as important than the durations of other categories in the speech acoustic inventory.
\item The appearance of position dependent phones among the top-20 indicates that duration of phones appearing in different parts of a word carry discriminating information on children growth.
\end{enumerate}

\section{Conclusion \& Future Work}\label{sec:conclusion}
In this work, we investigate features solely based on phone durations in speech (i.e., acoustic realizations of phonemes) for the task of speaker age estimation in children speech.
Phoneme occupancy distribution is derived by forced-aligning manual transcripts with speech signal.
Statistical functionals describing the distributions are extracted for each phone which constitutes the features.
A double layer stacking regressor architecture is employed with a meta estimator operating on top of multiple base estimators, each trained on statistical functional features corresponding to each phoneme.
The results suggests that phone durations contain critical developmental information helpful in predicting speaker age among children.
The results indicate that a speaker's age among children can be effectively predicted by looking only at the temporal variations in speech signal.
The best performing phone duration model yields mean absolute error of 1.62 and a correlation of 0.76.
The estimation of speaker age among children is associated with high error among young and older children, while yielding minimum estimation error among children studying among 2nd grade to 8th grade.
We find that aggregated phone durations of non-silence phones is the most important feature.
Among the other phonemes, particularly /T\_I/, /HH\_B/ and /K\_I/ play important role.
We also find that inter-speech silence duration also play an important role in predicting child speaker age.  Subsequent experiments on additional speech corpora, ADOS-Mod3, comprising speech data from children with ASD/ADHD diagnosis, further underscores the robustness of the phone duration features.

In the future, we plan to combine the phone duration features along with other speech based features including spectral features such as MFCC and voice quality features such as jitter, shimmer to explore complementary information for improved age estimation.
Additionally, combination of phone duration features and unsupervised total variability modeling based i-vectors or the supervised features derived through DNNs such as x-vectors can potentially complement and improve performance especially under low data scenarios.
We would also like to explore scenarios when manual speech transcripts are unavailable and thus alignments derived from an ASR is the only option, i.e., exploring the effect of automated transcriptions on the performance of speaker age estimation.

\bibliographystyle{IEEEtranN}
\bibliography{refs}

\end{document}